# Design of an Analog Memory Cell
# in 0.25 micron CMOS process

Bachelor of Technology (Honours)

in

Electronics and Electrical Communication Engineering

by

**Paramita Barai**

under the guidance of

**Dr. A. S. Dhar**

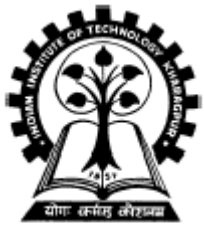

Department of Electronics and Electrical Communication

Engineering

Indian Institute of Technology

**KHARAGPUR**

**2002**

# *Certificate*

This is to certify that this thesis titled *"Design of an Analog Memory Cell in 0.25 micron CMOS process"* submitted by **Paramita Barai** to the Department of Electronics and Electrical Communication Engineering, Indian Institute of Technology, Kharagpur, in partial fulfillment of the requirements for the award of Bachelor of Technology (Honors) in 2002 is an authentic record of the work carried out by Paramita Barai under my guidance and supervision.

In my opinion, this work fulfills the requirements for which it has been submitted. This thesis has not been submitted to any other university / institution for any degree / diploma.

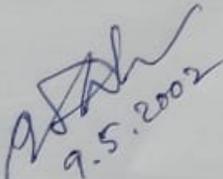

Dr A. S. Dhar

# Abstract

CMOS VLSI technology is the most dominant integration methodology prevailing in the world today. Various signal-processing blocks are made using analog or digital design techniques in MOS VLSI.

An important component is the Memory unit used to store data. In the project a memory cell has been built up using analog design method. A capacitor is used as the basic storage device. The main idea behind analog memory is that the analog value of the charge or voltage stored in the capacitor is the data stored. So the dielectric quality of the capacitor becomes important here to determine how effectively it can store some charge. Analog memory is a trade off between hardware cost, chip area and accuracy or quality of storage.

The circuit of analog memory cell was developed starting from the idea that required voltage will be stored in a capacitor and MOS transistors were used as switches. A given technology of integration was used and hence the dielectric property of the capacitor was fixed. By suitable circuit configuration the analog voltage value was written to the capacitor, read out when required and the charge loss was also refreshed.

The results obtained are as given in the thesis.

# Contents

# List of Figures

# List of Tables

# *Acknowledgements*

I am grateful to my supervisor Prof A. S. Dhar, for his encouragement and support throughout the project. I am indebted to him for providing helpful hints and suggestions at every stage of my work. I have gained immeasurably from his vast knowledge and experience.

I would like to thank my classmate Rajarshi Mukherjee in helping me regarding working with the software. He has helped me with learning the software tools that are used for design of integrated circuits and also have provided unhindered assistance in the understanding of the design process.

I would like to thank specially the IIT Foundation for setting up the Advanced VLSI Design Laboratory, which made my project possible. Special thanks to Dr. Bijoy G. Chatterjee of National Semiconductor for his help to the laboratory right from conception of the idea to the present state. I would also like to thank the employees at National Semiconductor's India Design Centre, Bangalore for their kind help in the setting up of the software environment.

I would also like to thank Samiran – da and Uttam – da for their service and help in working in the VLSI laboratory.

*Paramita Barai*
*09/05/02*

Paramita Barai

# Chapter 1

# Introduction

The Very Large Scale Integration ( VLSI ) technology today has made possible integration of millions of transistors on single Integrated Circuit ( IC ) chip. Modern VLSI has attained features in sub micron, e.g. 0.25-micron technology is very popular these days. The search is towards bringing more and more transistors in single chip by going more towards deep sub micron technology. Now the art is system – on – chip or integrating a whole system on a single IC chip that comprises of several analog and digital functions.

The VLSI design can be classified into three categories depending on the way of designing: Analog design, Digital design and Mixed Signal ( both analog and digital ) design. The regularity and granularity of digital circuits have made Digital designs automated to a good extent, with the help of CAD tools, given a behavioral description of the function desired. Whereas for analog circuit design a more hands on design approach is required. For mixed signal design a combination of both the ways is followed.

Various signal processing blocks are made using digital as well as analog design techniques. An important component is a memory unit, which has been designed using analog technique in the project. The basic idea behind the designed analog memory is that the analog value of charge stored in a capacitor determines the value that the memory stores.

# 1.1 Analog Integrated Circuit Design

Integrated circuit designing method is different for analog and digital designs. It may be useful to review definitions of analog and digital signals. A *signal* is an information of physical world expressible in quantitative terms like voltage, current, charge etc and detectable with easy means. *Analog Signal* is one which is defined continuously over time and amplitude / value; i.e. analog signal can have any amplitude at any time. *Digital signal* is defined at discrete values of time only and can have some finite distinct values of amplitude only.

At this point it will be significant to mention some of the differences between integrated and discrete analog circuit design. Discrete analog circuits consist of various active and passive components that do not share the same substrate. On the other hand, in an integrated circuit chip the various components of circuit share the same substrate. An advantage of this feature is, component parameter matching can be done more easily and this is used as a tool for design. At the same time question of isolation of devices sharing the same substrate becomes more significant and important; and proper care must be taken towards this. In integrated design the geometry of the active devices and passive components are under the control of designer; and this provides a greater degree of freedom in design as compared to discrete design. A discrete design is tested by wiring the circuit on a breadboard. Whereas an integrated designer must turn towards computer simulation by CAD tools to test the circuit performance. In integrated design there is always a limited range of components compatible with the design technology being used, as compared to discrete design where wider ranges of choices are available.

The task of designing an analog integrated circuit on chip includes several steps. The following flowchart gives a general sequence of steps followed in analog circuit design.

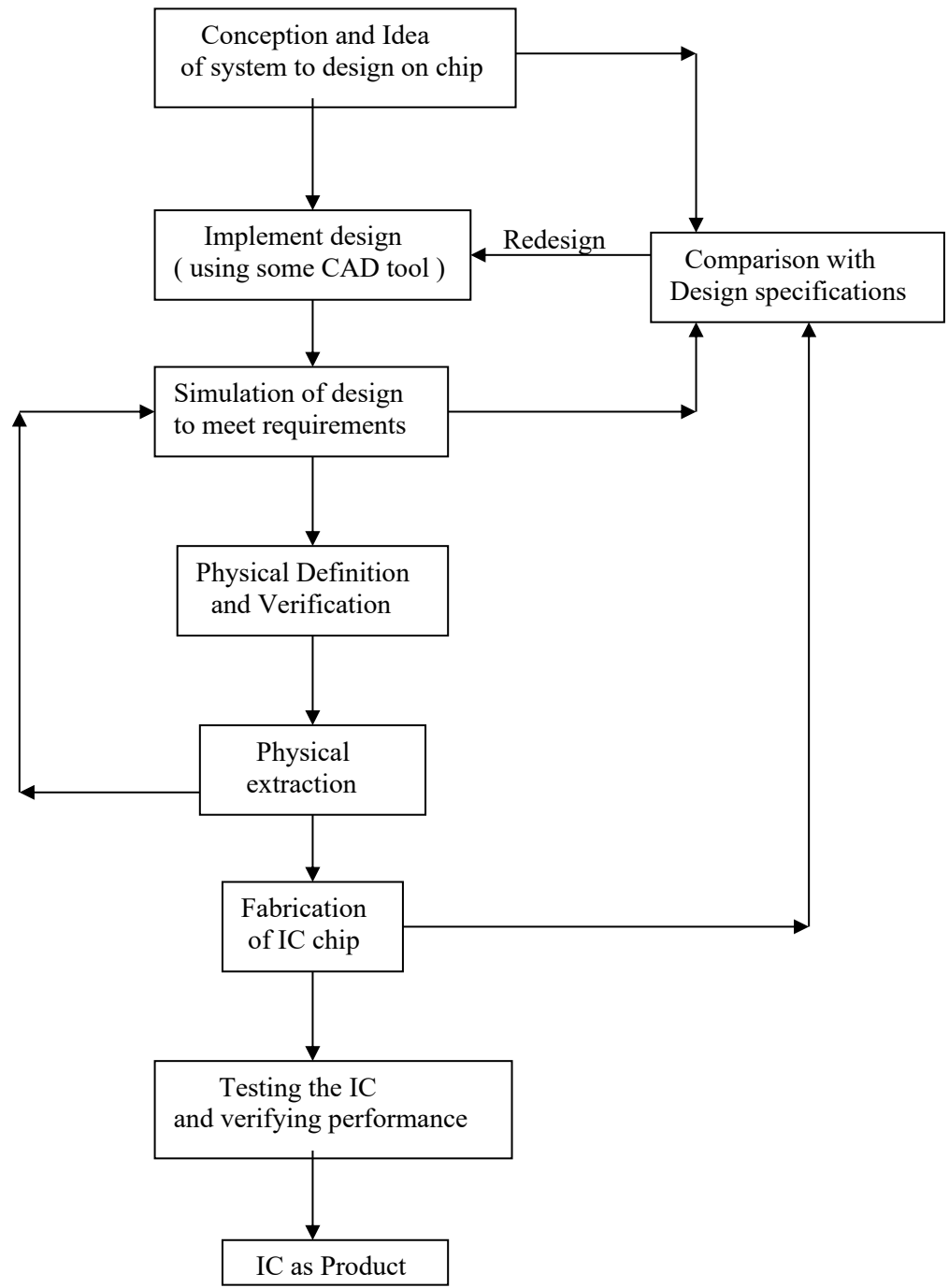

Fig 1.1     Flowchart of analog integrated circuit development

In words the flowchart can be explained as following. Firstly, the functions and specifications of the system to be fabricated are to be defined. These steps are very important as they determine the capability and performance of design. Next the circuit is to be implemented with help of some Computer Aided Design ( CAD ) tool. Then the performance of the circuit is to be simulated and seeing the results finally device parameters and orientations are to be adjusted to get required performance. After this layout of the finalized circuit is to be made. Then after extracting predictable and measurable parasitics from layout, simulation of circuit is done again. The whole process of extracting parasitics and re - simulating can be iterated several times to get better results regarding performance of circuit. Then from the final layout, the IC chip is fabricated. It is then tested and compared with design specifications to verify that Product is functional properly.

For analog design the above process is more difficult as compared to digital design; as the simulation of analog circuits is seldom very accurate. Often it happens that even after fabricating an analog chip taking all kinds of preventive measures; it does not work.

# 1.2    CMOS Technology

Nowadays most popular Integrated Circuit Design Technology are: Bipolar and Metal Oxide Semiconductor ( MOS ). The classification of various groups are as following.

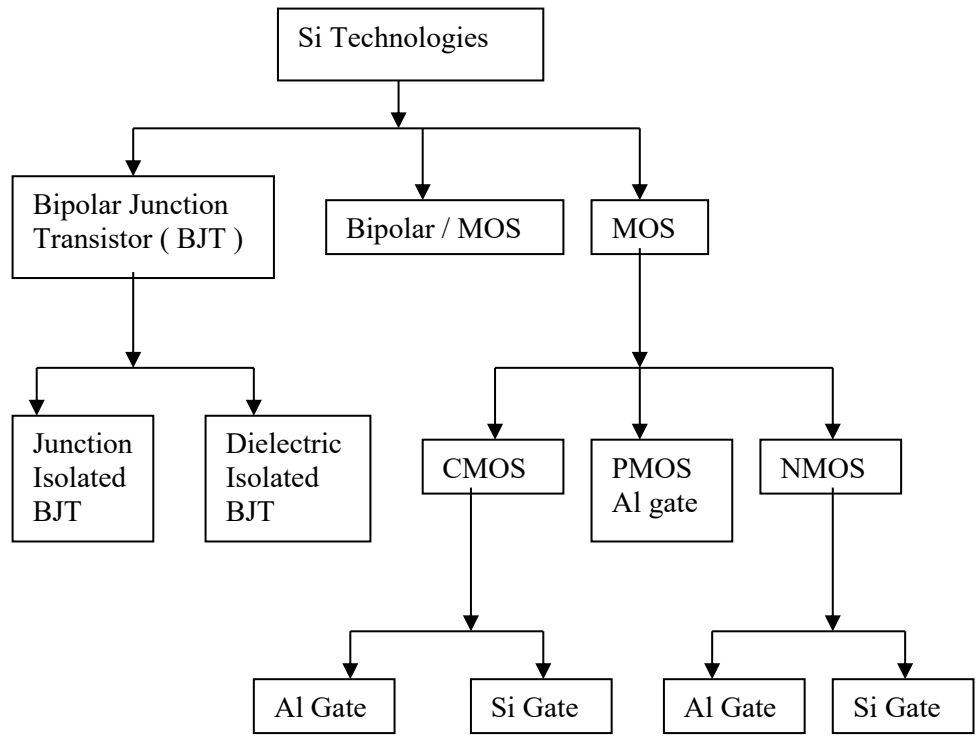

Fig 1.2 : Categories of Silicon Technology

When design of system on chip is the question then CMOS technology gives more advantages compared to the others. So analog design using CMOS is very popular today. Bipolar Complementary Metal Oxide Semiconductor (BiCMOS) technology is used often when a high degree of performance is required. BiCMOS is much costlier then CMOS design. Hence still the most prevalent design technique is CMOS.

## 1.3  Comparison of CMOS and Bipolar analog IC

- Transconductance of bipolar transistors is very high. So with bipolar it is possible to have circuit of very high gain and high gain bandwidth. CMOS designs are limited to low gains.

- Bipolar circuits use linear loads. With an aim to minimize the area as much as possible linear loads are not used in CMOS circuits. Therefore the signal handling capability of CMOS circuits is lower compared to bipolar.

- Bipolar transistors can work at a lower voltage supply than CMOS. But by modern technology $V_T$ of CMOS have decreased to ~ 0.5 volts. This has brought in several advantages.

- In CMOS design a higher degree of matching can be attained using special layout techniques.

- Basically CMOS technology is more popular because of the main advantage of high packaging density; and the strengths of CMOS has taken power over it's weaknesses making it the most popular VLSI technology today.

## 1.4  Analog Memory using CMOS technology

In the project a circuit for an analog memory cell was developed. A definite CMOS technology was used for it, the *0.25 micron CMOS process*. A capacitor was used to store the analog voltage as the charge across it. MOS transistors were used as switches to control operations like writing the voltage value in the capacitor, reading the voltage from it. The charge loss from the capacitor was compensated by refreshing it at regular intervals after amplifying the dropped voltage by an op amp.

# 1.5  Organization of the Thesis

The thesis describes the design of an analog memory cell. Chapter 2 gives the initial motivation for going for analog kind of design and gives a flavor of some previous works done in this field. The actual design of the analog memory unit is given in Chapters 3 and 4. Chapter 3 deals with the preliminary circuit designing where voltage to be stored is written to a capacitor and whenever required voltage is read from it. No provision is yet taken to make up for the charge loss through leakage paths. Hence the voltage read is not the correct value that was initially written in the capacitor. Chapter 4 gives the complete self-compensating analog memory circuit. Here the dropped voltage is amplified and refreshed at regular intervals so that voltage across the capacitor is constant for a pre defined time interval within certain tolerance limits. Chapter 5 gives the conclusions drawn and scopes for future work.

Appendix A gives a brief overview of CMOS models used in design. These models are not very accurate, but give insight into the design process. Appendix B gives a brief description of the Cadence IC 445 set of CAD tools that were used for the design and simulation of the circuits from beginning to end.

# Chapter 2

# Background

## 2.1  Reason for choosing Analog design

Today's world is going more and more towards digital kind of processing. This is because of several advantages of digital design like accuracy, precision, greater noise margin, greater flexibility etc.

A typical Digital Signal Processing Block is as shown following.

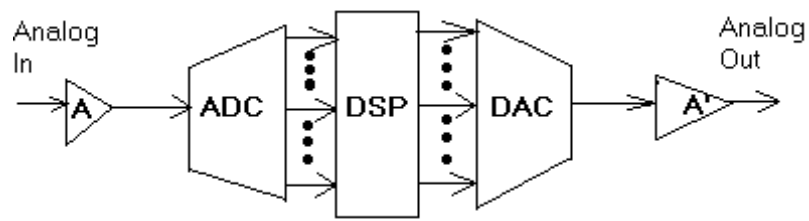

Figure 2.1    Typical Digital Processing Block

But in digital designs hardware cost and speed brings the limitation. If we try to manufacture digital signal processors operating at very high speeds ( several GHz range ) then cost of system increases tremendously. Also the digital processors have some high-speed limitation as propagation delays of some basic gates come into picture and don't allow operation of the processors beyond some highest speed.

Hence in some specific type of utilities analog design may prove to be much more useful than digital design. These are those kinds of applications where both low cost and somewhat high speed are required together.

One of the most important drawbacks of analog design is it's noise immunity. The analog values of voltages and currents are much more prone to get changed by the slightest noise disturbance from outside. Hence inaccuracy creeps up in circuit performance very easily. But now, by some modern integration technologies it has been possible to reduce the noise affecting the components of a chip.

By today's advanced VLSI technology amount of noise in IC's have decreased. In earlier days typical voltage supply was     + / - 5 volts. So in Digital designs noise margin was up to 2.5 volts as the actual noise was comparable to that. In today's technology generally 0.9 volts / 1.1 volts power supply is used which means noise margin is reduced to 0.45 volts / 0.55 volts. This has been possible because, by special integration methods noise in digital designs have actually reduced. The design principles used in digital domain which gives reduced noise, similar ones are used in analog design so that noise is less. This reduced noise is used to make analog design more effective.

The block diagram of a typical analog Signal-processing block is as shown next.

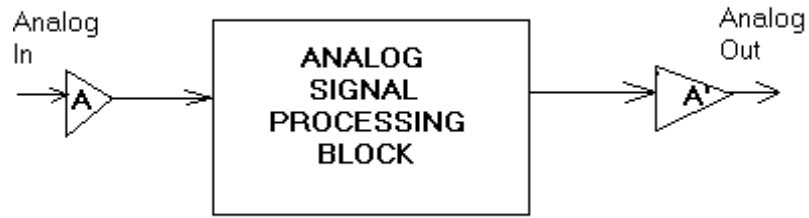

Figure 2.2    Analog Signal Processing Block

Another advantage of analog design is clear from the above figure. Here ADC ( Analog to Digital Converter ) and DAC ( Digital to Analog Converter ) blocks are not required; this saves a good amount of hardware and hence makes cost efficient design.

Analog kind of design basically brings a trade off between quality of performance, accuracy, hardware cost and speed. An example of such analog design is Analog Memory.

## 2.2  Analog Sampled System

Now analog design means design of an analog sampled system. Here processing is discrete in time domain but continuous in amplitude.

Example of IMAGE PROCESSING

Considering an image of size 256 pixels * 256 pixels.

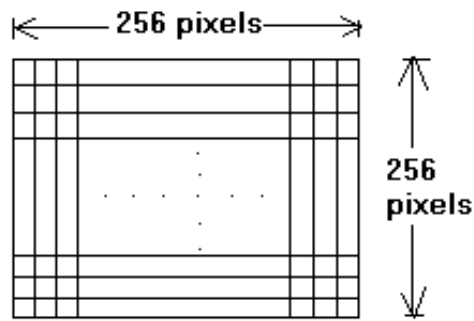

Figure 2.3    A 256 pixel * 256 pixel Image

In digital kind of memory i.e. in digital SRAM or DRAM image data is expressed digitally and stored as bits, 1 or 0. If we consider word length of 8 bits, then each analog value or pixel is converted to a digital word 8 bits long, where voltage resolution is 1/256 ( since $2^8 = 256$ ). In normal kind of SRAM / DRAM design, to store one word, 8 capacitors or 8 MOS transistors are required; each to store value of one bit; 1 / 0.

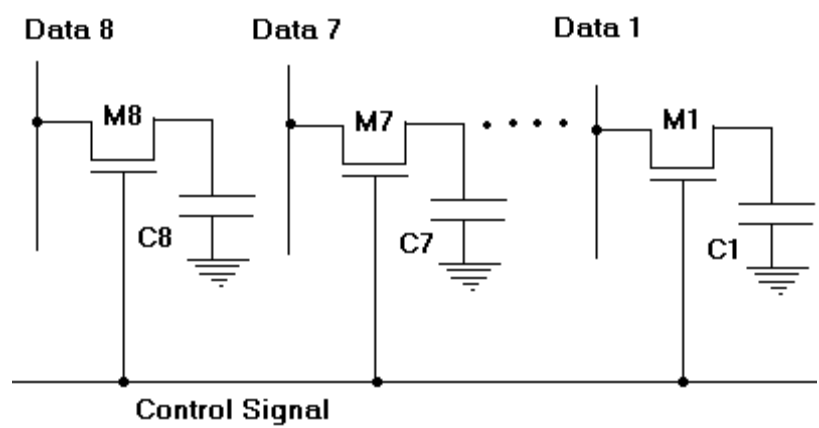

Figure 2.4    DRAM structure for storing 8 bit digital word

The digital kind of storage is discrete in both time and amplitude. The whole image is divided into pixels of finite number that makes it discrete in time domain. And each pixel can be one of the 256 ( = $2^8$ ) distinct levels. That makes it discrete in amplitude.

Assuming each pixel is one word ie. 8 bits long, if the image is stored in digital memory then number of capacitors required to store the image will be 256*256*8 = 524288.

On the other hand in analog kind of storage only 1 capacitor is required to store one word or one pixel.

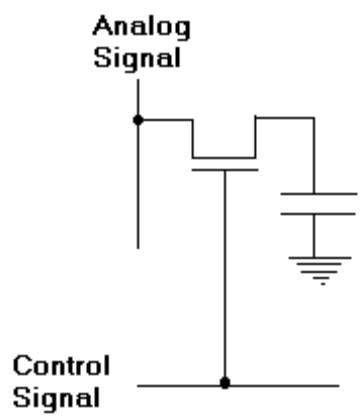

Figure 2.5    Analog memory structure

Here the analog value of the charge stored in the capacitor represents the Grey Level of the pixel in the image. The image is divided into 256 pixels * 256 pixels; so it is still discrete in time. Only the processing has become continuous in amplitude as each pixel can attain any of the infinite values possible between minimum and maximum limits of pixel grey level.

The basic principle is that the analog value of the charge stored in the capacitor represents the value stored in the memory cell.

Successful analog memory design depends entirely on the dielectric quality of the capacitor used for storing signal value. If the capacitor is of very high quality i.e. it's leakage current is very small; then the charge lost from it by leakage will be negligible for a small amount of time. Hence in that amount of time the charge stored can be assumed to be practically constant and the voltage read will give the correct value of voltage stored.

The analog memory operation can be classified into two parts depending on the processing and hardware complexity.

## 1.     Short – time storage:

The pixel value is used only once within very short time of storing the value; so that in the small time interval leakage of the capacitor is negligible. Then there is no need of refreshing the charge in the capacitor; and the value read is the correct value stored.

## 2.     Long – time storage:

If the pixel value is needed to be stored for a longer time then leakage of capacitor becomes significant and charge loses considerably. Then the charge needs to be refreshed and this aspect of design is the challenging job. As exact analog values are important here; looking at the present value of charge an estimate must be made about the original value stored. The dielectric quality of the capacitor determines the refreshing interval.

In the project the aim was to design analog memory unit for long time storage, so the charge loss was significant and refreshing was required. Further a given CMOS technology was used which fixed the quality of components. Hence dielectric quality of capacitor was fixed and by proper circuit configuration charge loss was refreshed.

## 2.3  Review of some previous works

## 2.3.1 Analog Memory Elements

The simplest structure for storing charge is the DRAM style cell shown in Fig 2.6. However the leakage current of the source of the switch transistor limits the charge retention time, to up to a few seconds for digital DRAM storage cells. The acceptable retention time for an analog application obviously depends on the resolution required. Considering the resolution, the acceptable retention time drops from that for DRAM cells by two orders of magnitude (for 8 bit resolution by about 1/256). This circuit is useful only for very short term storage, such as in small imagers with fast frame rates.

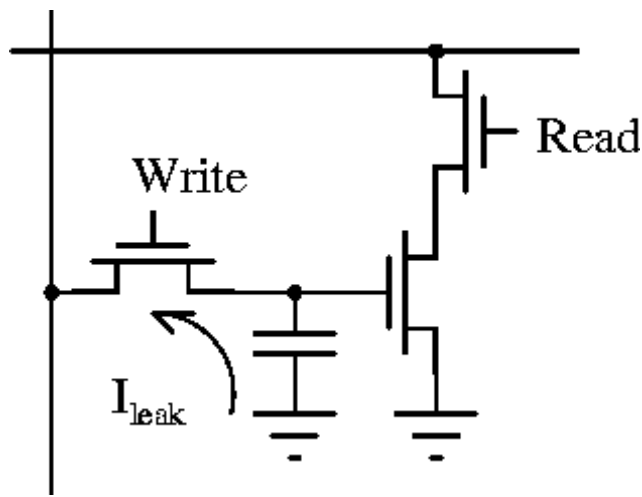

Figure 2.6:  A DRAM style memory for storing analog charge.

The storage capability of the cell can be improved by using several techniques, such as differential storage and leakage reduction. In differential storage technique the original signal is translated into a differential signal and stored on two similar storage devices. As the leakage reduces the charge at both nodes almost equally, the difference still remains the same. This method can increase the storage time by several times. A drawback of this technique is the additional area consumed by single ended-to-differential translation and the extra capacitance.

In leakage reduction techniques the leakage of the source/drain diffusion of the switching transistor at the storage node is reduced by setting the voltage across the anode and cathode of the source diffusion-well diode to zero, as shown in Figure 2.7. Using this circuit storage times of up to several seconds in normal conditions can be achieved.

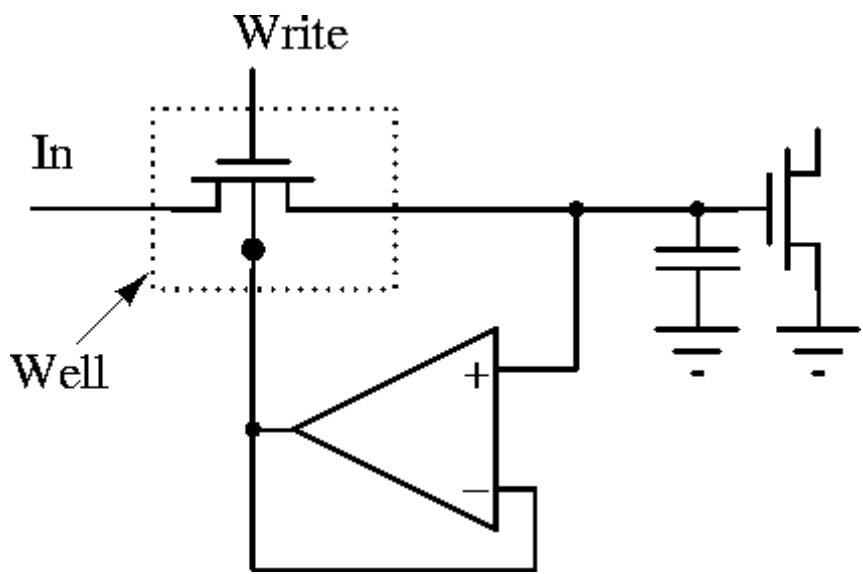

Figure 2.7: Circuit for reducing the leakage.

Floating gate structures have long been used in EPROM devices. Their storage capability for analog signals has also been used in many implementations of analog systems. Despite the long term storage achieved with floating gate structures (in the order of several years), the accuracy, programming, and reprogramming issues of these devices remain to be challenging. Floating gate devices can be found either in special processes, where thin-gate devices are available for low voltage programming, or in standard processes, where the gate of a normal transistor is left floating. The floating gate devices in standard processes require high programming voltages, which might exceed the breakdown voltages of different junctions in the process, or they may need accelerated mechanisms by exposing the chip to UV light. The accuracy achieved using floating gate devices is around 6 to 8 bits.

## 2.3.2  Simoni et al.'s Optical Sensor and Analog Memory Chip with Change Detection

In [7],[8] Simoni et al. describe an analog memory with peripheral circuitry for difference detection. Each pixel comprises a photodiode and a storage capacitance. When a particular pixel is selected, first the previous pixel value which is stored in the capacitor is read out, then the present value is read from the photodiode. A circuit computes the difference between these two values. The architecture and the circuits are based on the switched-capacitor technique. The schematic of a cell is shown in Figure 2.8. It should be noted that the change detection capability of this chip is not based on algorithmic models.

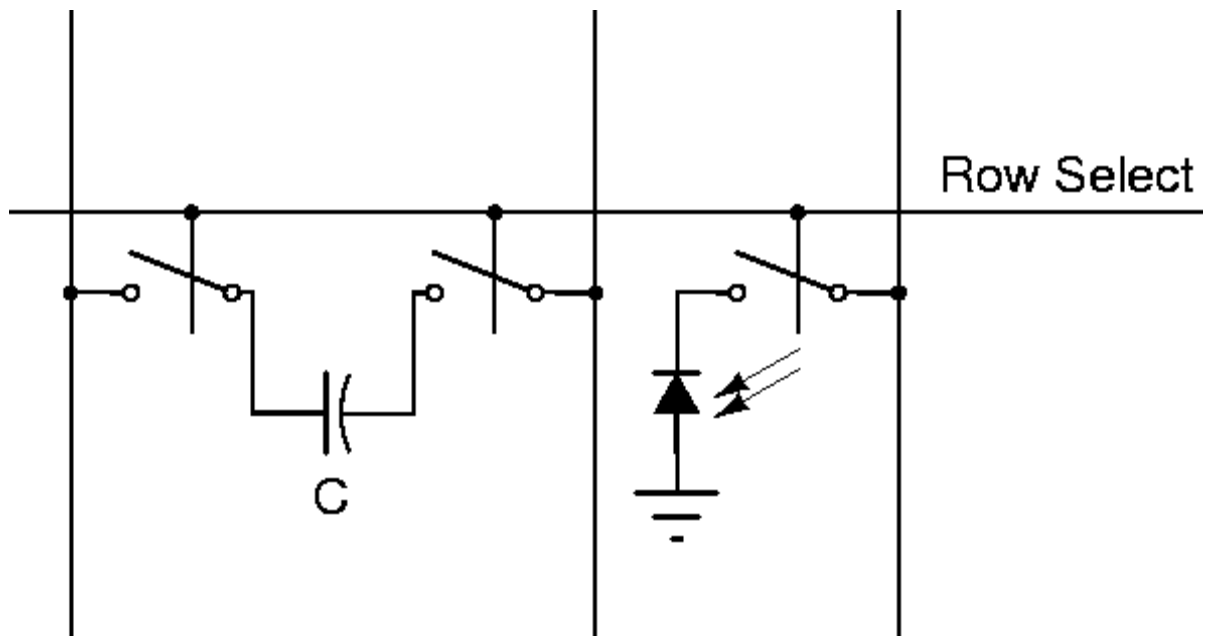

Figure 2.8: Schematic of a cell of Simoni et al.'s motion detection chip.

# Chapter 3

# Design of the basic Memory Cell: with writing and reading facility

Memory cells are used to store some value of voltage, current or charge that is later read out for signal processing purposes. In Analog Memory the value is stored as charge or voltage in analog domain. This means that the actual analog value of the voltage stored determines the value that the memory contains.

The power supply used in the circuit simulations is 2.5 volts. Noteworthy at this point is, in the 0.25-micron technology the highest allowable power supply is 3.3 volts.

## System specifications :

- Analog Voltage is to be stored across capacitor till 120 milliseconds.
- Standard Frame rate is 50 frames per second but it must also be able to work with slow rates e.g. 8 frames per second.
- The voltage is to be refreshed after fixed time interval depending on the frame rate. This means whenever it is the time for a new frame to arrive, original value stored is to be refreshed, read out and after that new value written.
  ( Analysis was done assuming 25 frames per second or refreshing after every 40 milliseconds. )
- Power supply used is dual supply of + 2.5 volts and – 2.5 volts.
- Maximum input voltage to be stored is 2 volts.

# 3.1 Initial idea to store charge in a capacitor

## 3.1.1 Circuit

The most basic idea of an analog memory cell i.e. storing the charge in a capacitor in analog way is as shown in figure 3.1.

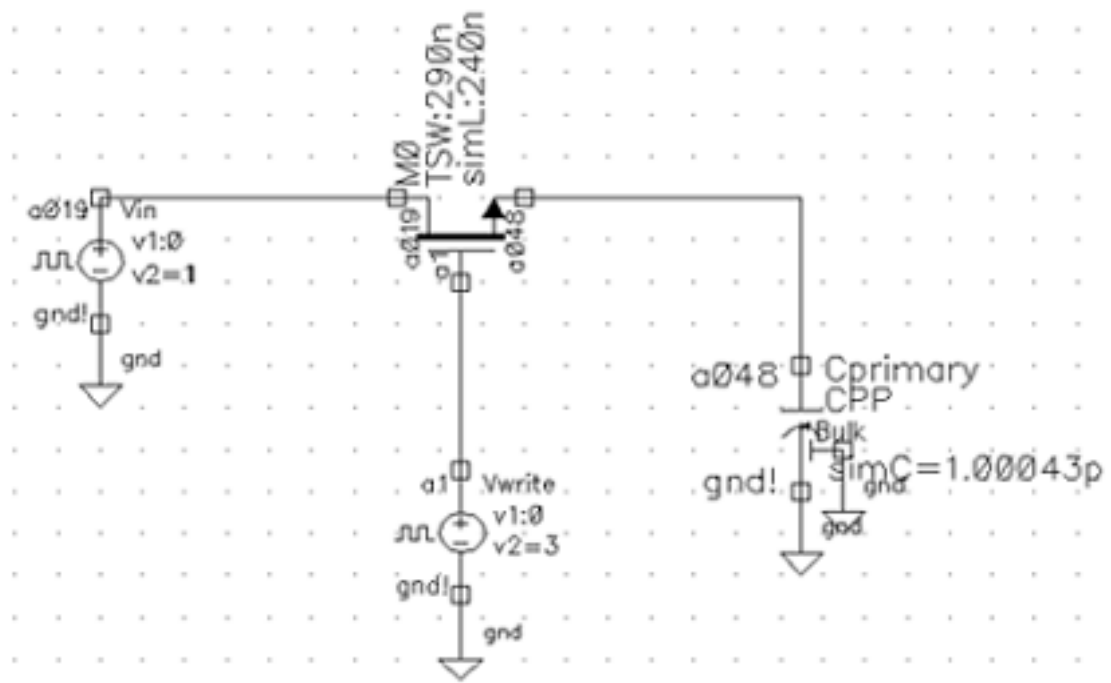

Figure 3.1 Main scheme of storing charge in a capacitor in analog domain

The NMOS $M_0$ acts as a switch. The input voltage ($V_{in}$) is given at drain of $M_0$. The switching voltage ($V_{write}$) is given to gate of $M_0$. The capacitor $C_{primary}$ stores the required voltage, $V_{in}$ as charge across it. Whenever $V_{write}$ is HIGH the MOS switch $M_0$ is ON i.e. allows voltage $V_{in}$ to reach capacitor $C_{primary}$. After $V_{write}$ is HIGH for some time the capacitor gets fully charged. Then $V_{write}$ is made LOW, i.e. MOS $M_0$ is switched OFF.

If the capacitor $C_{primary}$ were ideal, then when left by itself it would retain the charge stored across it. But because of leakage currents through various paths the charge gets leaked out gradually and hence voltage across $C_{primary}$ falls. The rate of fall is shown next in results.

## 3.1.2        Simulation and Results

All the results shown below and henceforth have been obtained by simulation in the CAD tool software *'ace445'*. The results are those of transient response i.e. plots of voltage or current at any node with respect to time.

It was found that $V_{write}$ must be HIGH to a minimum time of 40 nanoseconds. And $V_{in}$ must be present till a time greater than 40 nanoseconds, to write the signal value applied as $V_{in}$ to the memory successfully.

The voltages at various points of circuit are as shown in the next graphs.

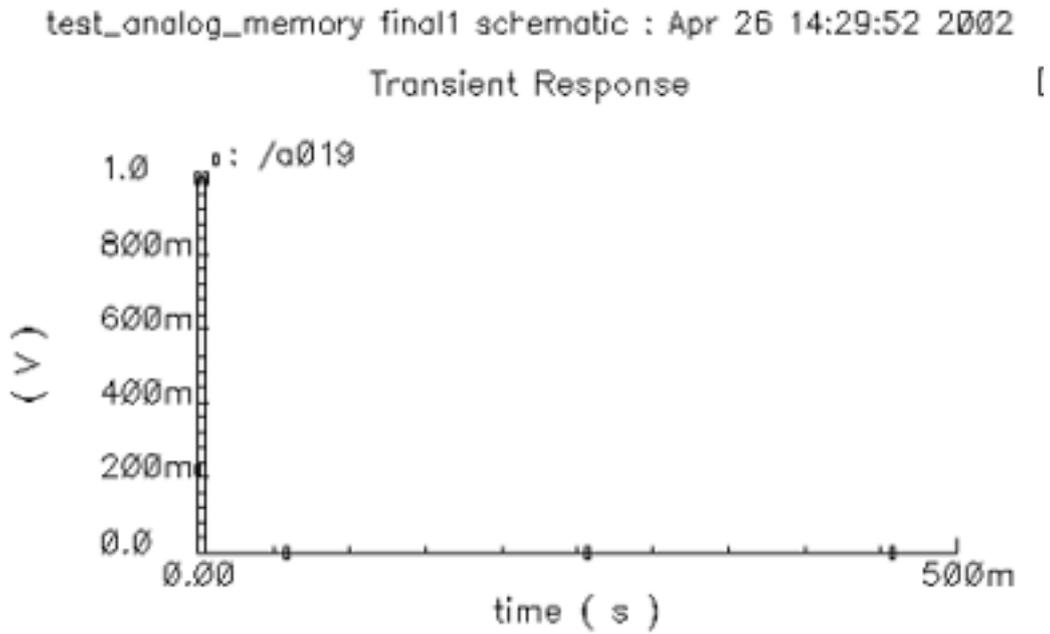

Figure 3.2    Signal Value or Input voltage ( $V_{in}$ ) to be stored

The above shows $V_{in}$ , which is the signal value to be stored in the capacitor. In simulation $V_{in}$ is applied as constant DC voltage of duration 42 nanoseconds.

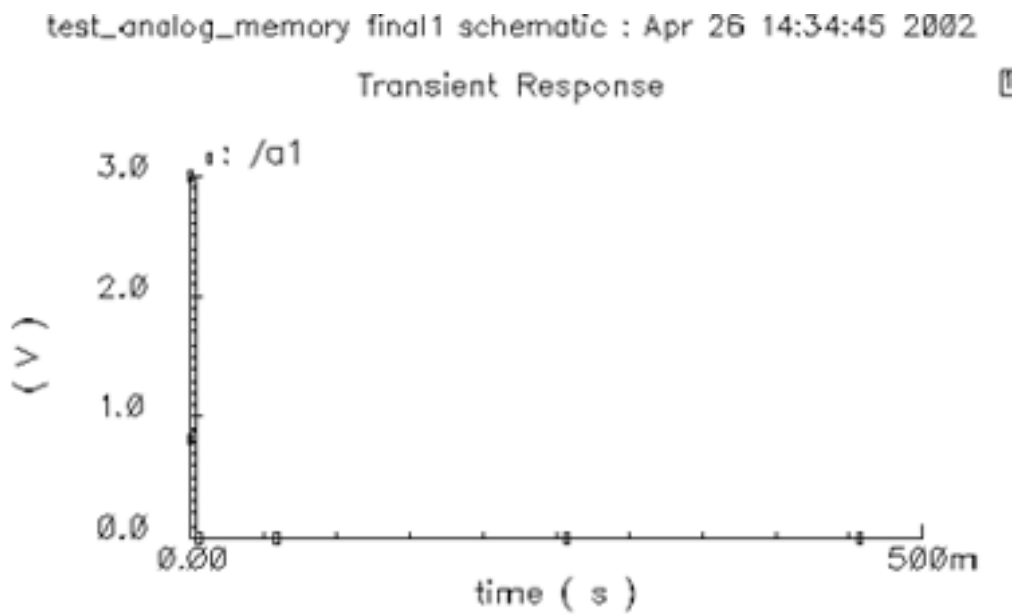

Figure 3.3    Write Control Signal ($V_{write}$ )

The previous figure shows $V_{write}$, which is the control signal used to switch ON the MOS switch, $M_0$. In simulation $V_{write}$ is applied as constant DC voltage of duration 40 nanoseconds.

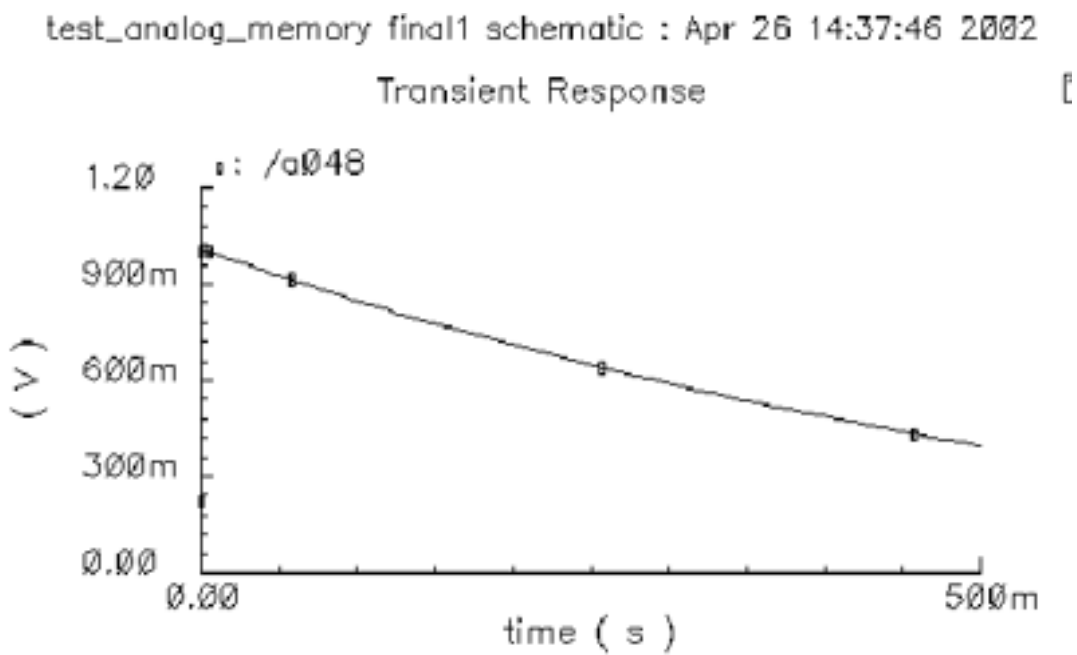

Figure 3.4    Voltage across $C_{primary}$

The graph above shows how the a voltage value ( of 1 volts in this plot ) applied initially across $C_{primary}$ , falls with time. This much charge loss is not desirable, so there must be some provision to refresh the charge stored in $C_{primary}$. Also to know the correct voltage stored at a considerably later time, voltage must be amplified with proper gain, before reading.

## 3.2 Introducing another capacitor for reading out stored voltage

### 3.2.1 Circuit

To read the voltage stored the mechanism as shown in following circuit is chosen.

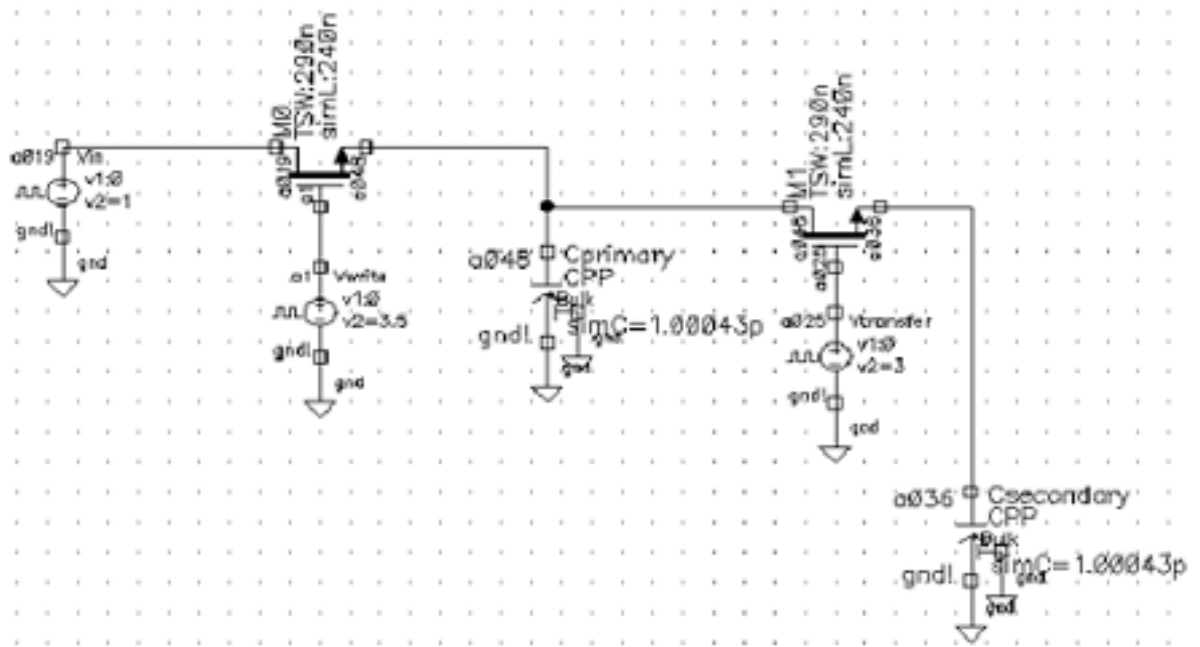

Figure 3.5    Circuit with secondary capacitor for memory read out

If we read the voltage stored directly from $C_{primary}$, then the charge across it falls very much and hence for refreshing the voltage after reading it, the gain required is very high. So another capacitor $C_{secondary}$ is introduced. $C_{secondary}$ is used to store the charge temporarily from where the voltage level is read out. The NMOS $M_1$ acts as the switch to transfer charge from $C_{primary}$ to $C_{secondary}$. When $V_{transfer}$ is HIGH , $M_1$ is ON and makes contact between the two capacitors. Hence charge sharing occurs between $C_{primary}$ and $C_{secondary}$. The voltage stored across $C_{primary}$ gets halved when charge is redistributed between $C_{primary}$ and $C_{secondary}$. So the gain required to refresh the initial voltage becomes doubled.

With this arrangement $C_{secondary}$ gets charged through leakage paths in the meantime i.e. when not required. So when $V_{transfer}$ is HIGH the charge between the two capacitors gets distributed among each other equally. This redistributed charge includes the leakage charge of $C_{secondary}$ also, which adds error to the actual charge stored. The following graphs give an estimate of how the leakage charge affects the true voltage stored across $C_{primary}$.

## 3.2.2    Simulation and Results

The voltage across the capacitors are as shown in the next graphs.

The first graph is for the case when $M_1$ is OFF or $V_{transfer} = 0$ throughout. So ideally no voltage should appear across $C_{secondary}$ as no charging path is provided through $M_1$. Here no charge sharing or distribution occurs between $C_{primary}$ and $C_{secondary}$.

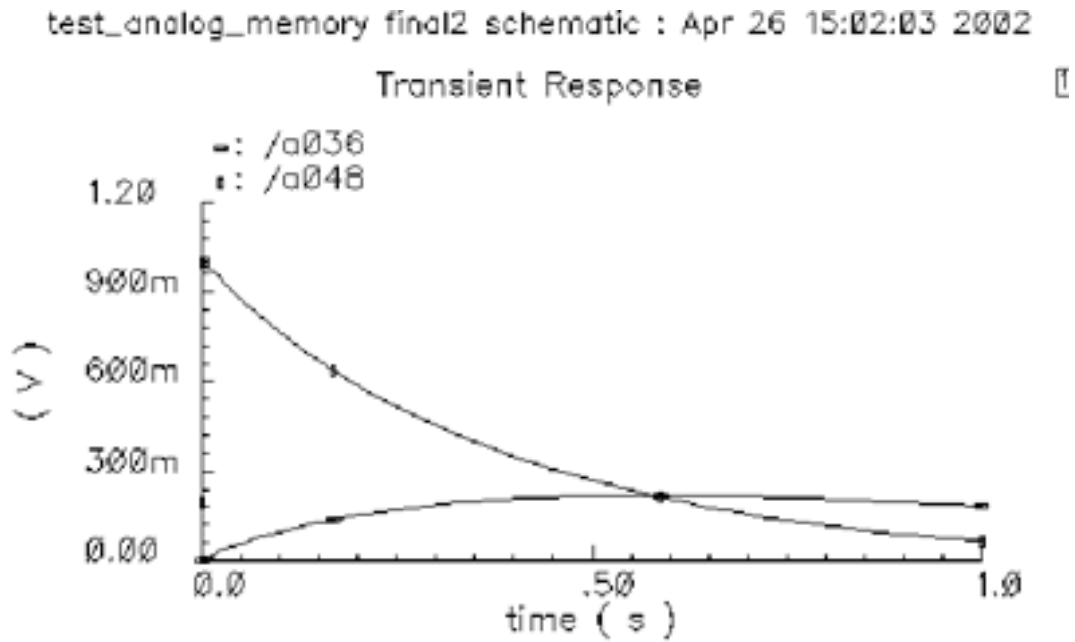

Figure 3.6    Voltages across primary and secondary capacitors

                          a048  :       voltage across primary capacitor

                          a036  :       voltage across secondary capacitor

The above graph implies that the secondary capacitor gets some leakage path ( other than through $M_1$ ) and charges up to voltage of few 100's of mV.

The next graphs are for the case when pulse train of amplitude 3 volts is applied as $V_{transfer}$. So whenever $V_{transfer}$ is HIGH charge sharing occurs between primary and secondary capacitors. And at those instances voltages of both capacitors become equal. When again $V_{transfer}$ becomes LOW, meaning no contact between the two capacitors each of them loses charge at own rate.

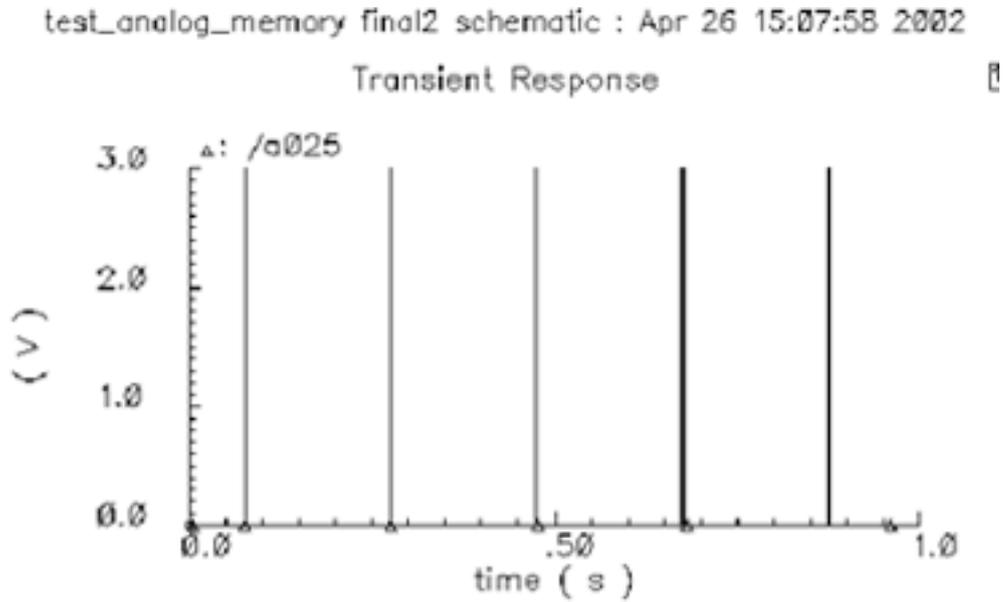

Figure 3.7    V_transfer, Signal controlling transfer of voltage between primary and secondary capacitors

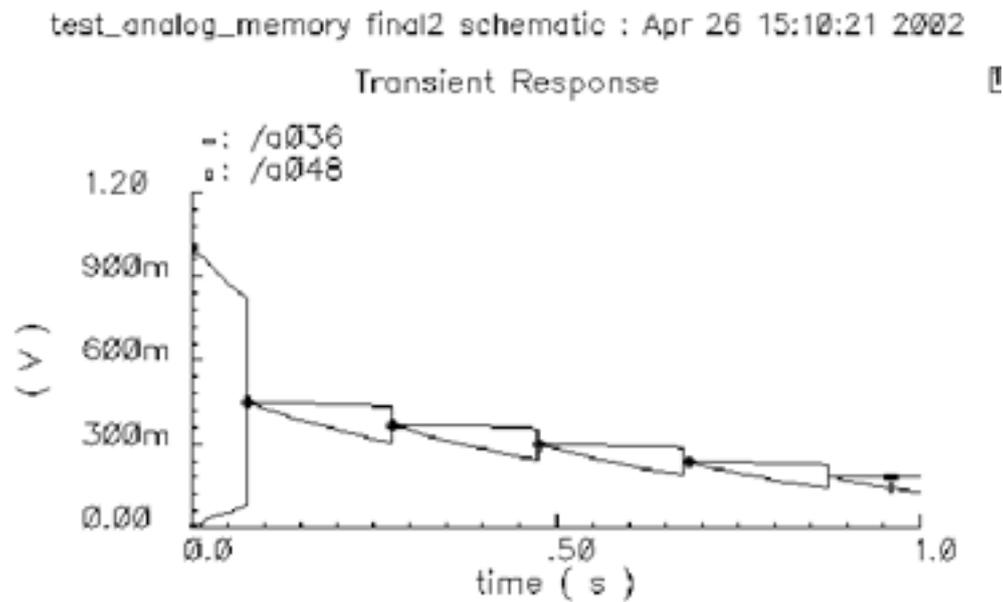

Figure 3.8    Charge sharing between primary and secondary capacitors

a036  :      voltage across $C_{secondary}$

a048  :      voltage across $C_{primary}$

Next figures 3.9 and 3.10 shows same thing as Fig. 3.8 only in two different plots for sake of understanding.

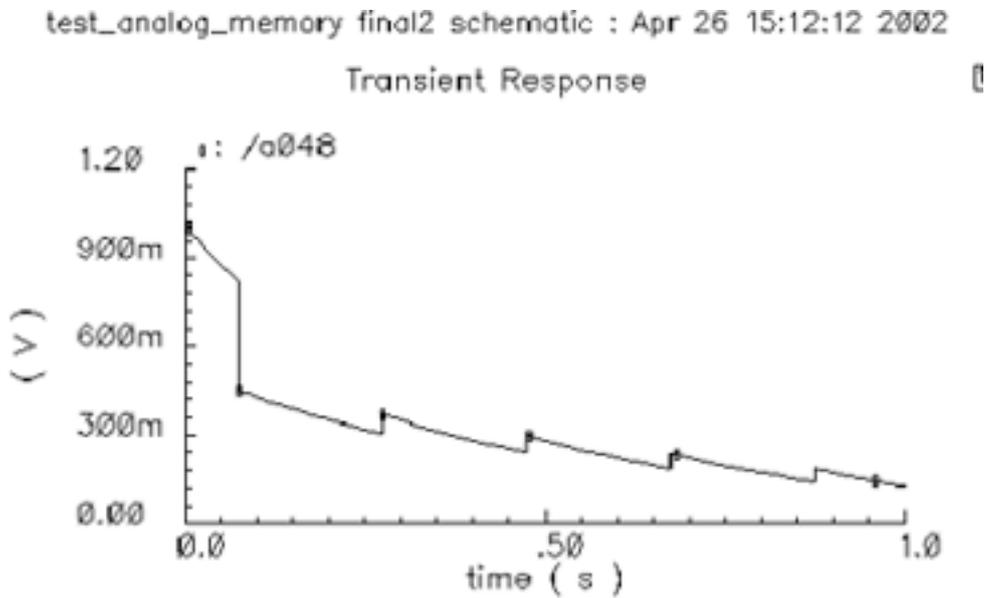

Figure 3.9    Voltage fall across primary capacitor

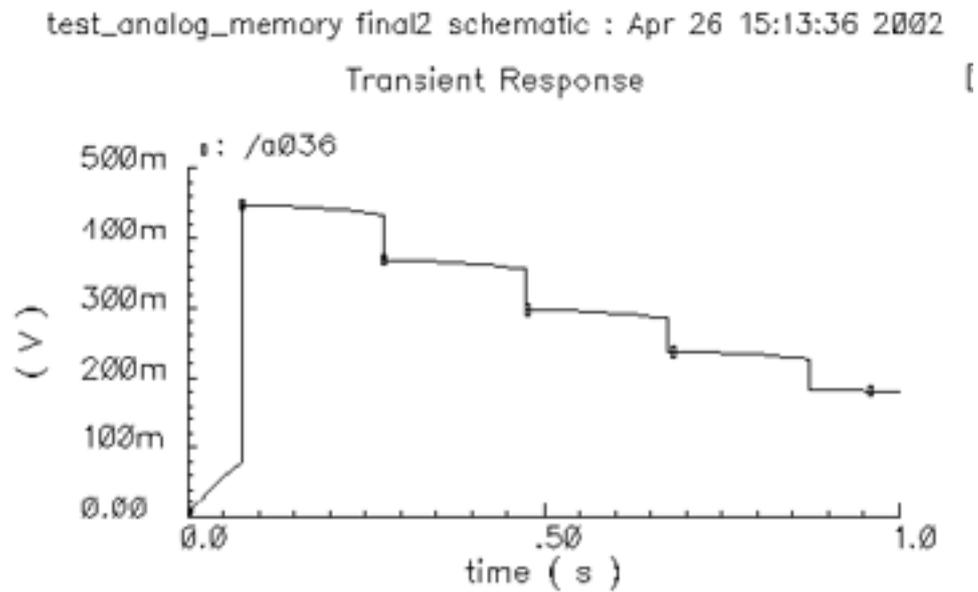

Figure 3.10  Voltage across secondary capacitor

So whenever $V_{transfer}$ is HIGH, charge sharing occurs between $C_{primary}$ and $C_{secondary}$; and the voltage across them becomes equal at those instances of time. After that voltages across the two capacitors fall at their own rates and again become equal when $V_{transfer}$ is HIGH.

The leakage charging of the secondary capacitor introduces error in the voltage stored. To prevent this leakage charging, $C_{secondary}$ must be kept discharged when not required.

## 3.3 Discharging the secondary capacitor to prevent it's leakage charging

### 3.3.1 Circuit

The circuit where the secondary capacitor is discharged when not required is as shown next.

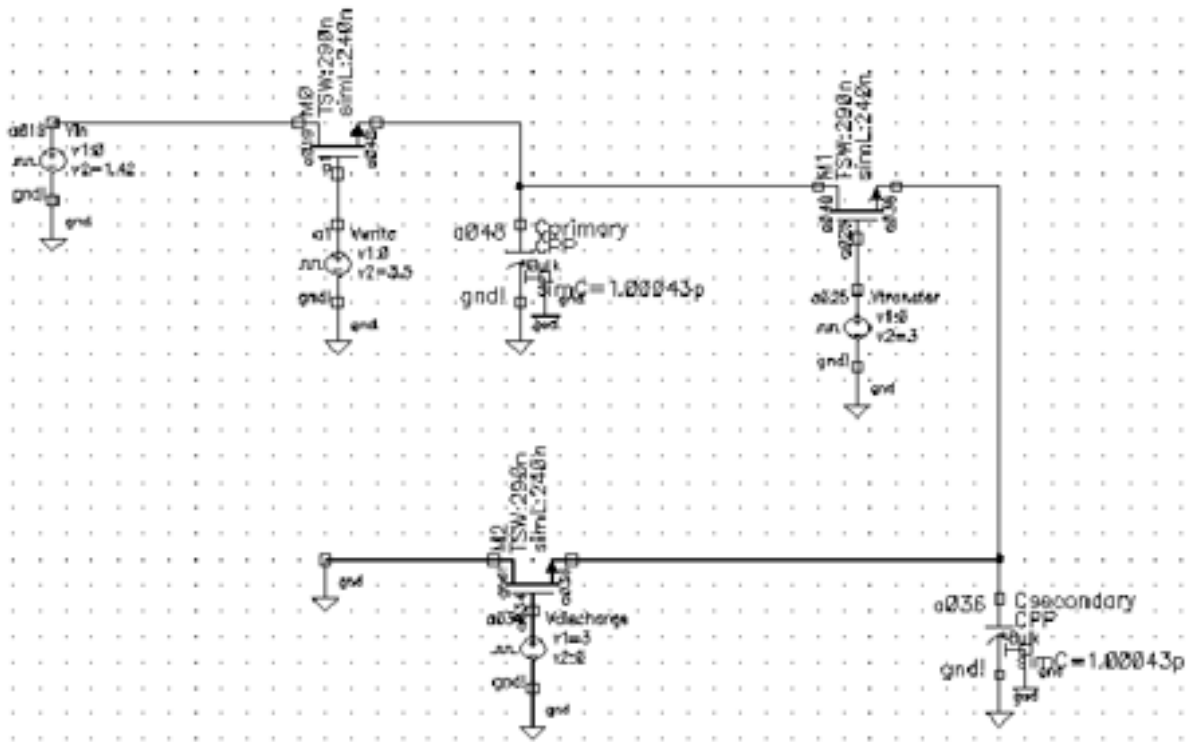

Figure 3.11  Circuit with discharging of secondary capacitor

Here NMOS $M_2$ acts as the discharging switch. Whenever $V_{discharge}$ is HIGH, $M_2$ is ON and charge of $C_{secondary}$ gets direct path to ground through $M_2$. So then $C_{secondary}$ gets discharged and voltage across it is zero. When charge sharing is wanted between $C_{primary}$ and $C_{secondary}$; $V_{discharge}$ is made zero so that $M_2$ is OFF and charge across $C_{secondary}$ gets no discharge path. Then by switching $M_1$ ON, by making $V_{transfer}$ HIGH; voltage is transferred from $C_{primary}$ to $C_{secondary}$. Soon after that voltage is read from $C_{secondary}$ and then it is again grounded by closing switch $M_2$.

## 3.3.2    Simulation and Results

The rate of fall of the charges across $C_{primary}$ and $C_{secondary}$ , their sharing of voltages when discharging of $C_{secondary}$ is taken into account is as in next graph.

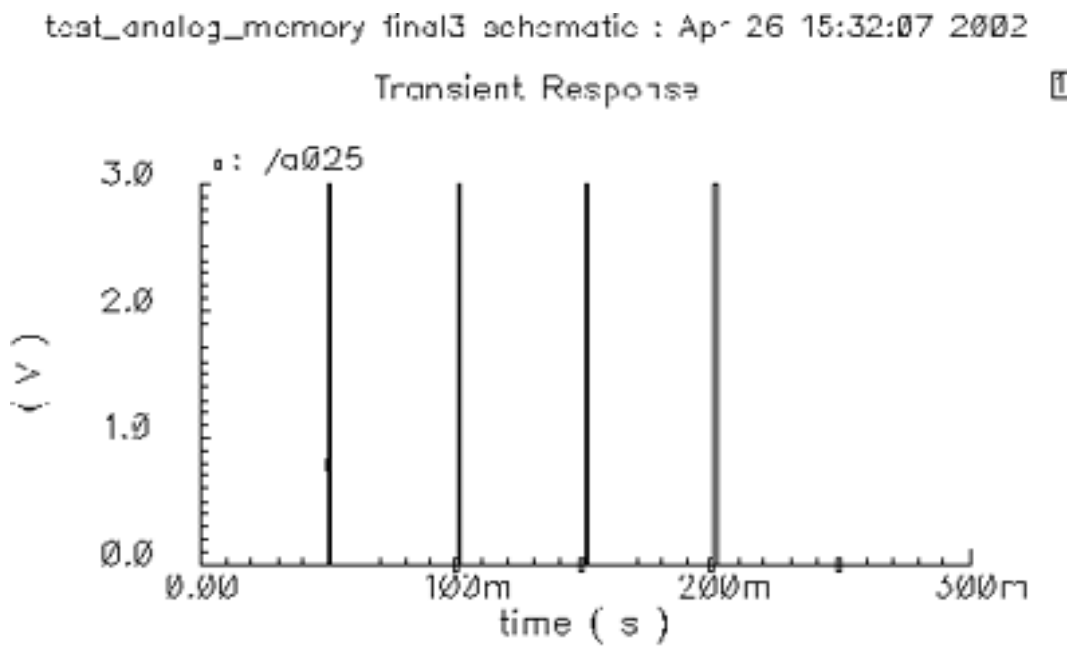

Figure 3.12  Signal $V_{transfer}$ applied

The above graph shows the time instances when sharing of charge between primary and secondary capacitor occurs.

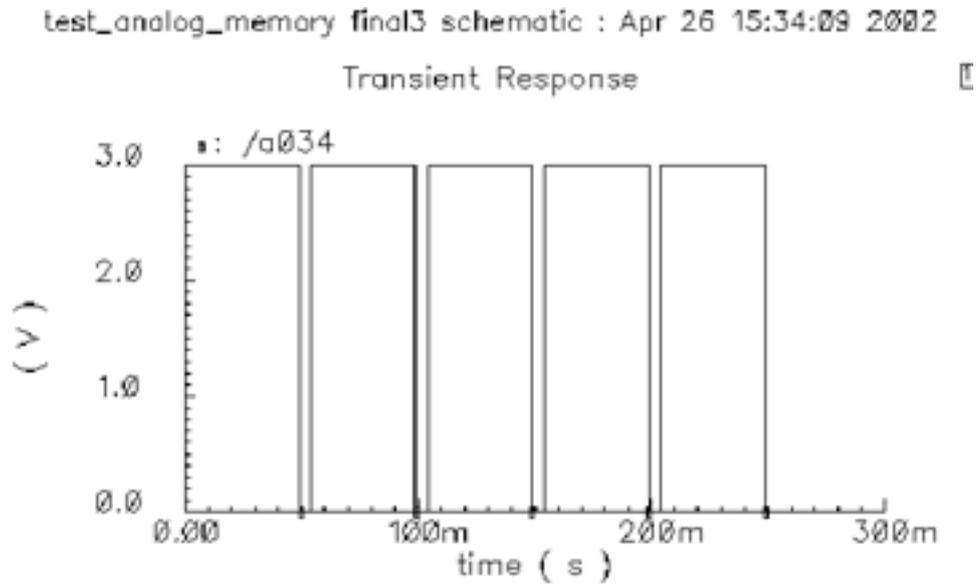

Figure 3.13  Signal  $V_{discharge}$  applied  to  discharge  $C_{secondary}$  when  not required

When $V_{transfer}$ is not HIGH, $C_{secondary}$ should be discharged and so $V_{discharge}$ is HIGH at those times.

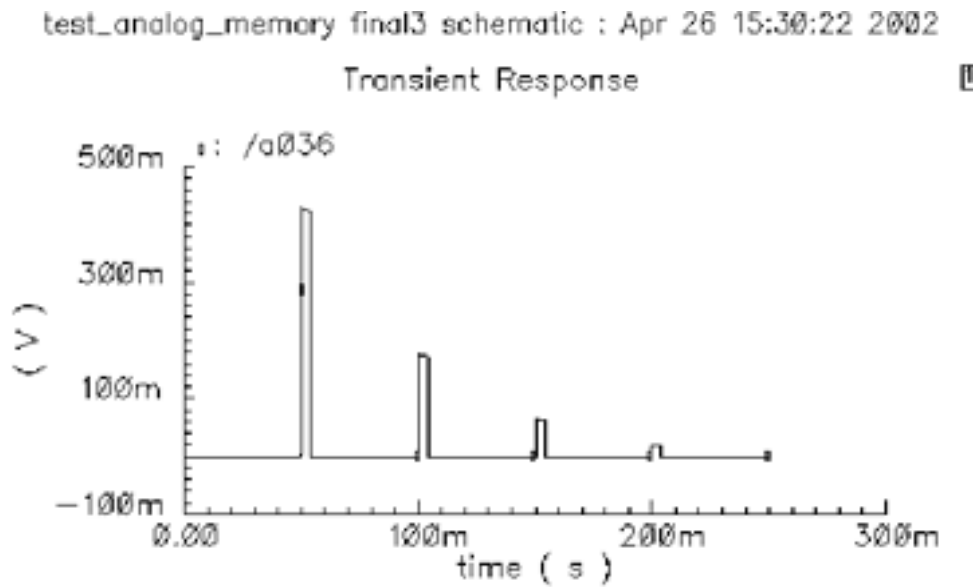

Figure 3.14  Voltage across $C_{secondary}$

When $V_{discharge}$ is LOW and $V_{transfer}$ is HIGH $C_{secondary}$ stores part of charge from primary capacitor and the voltage across it is shown in previous figure.

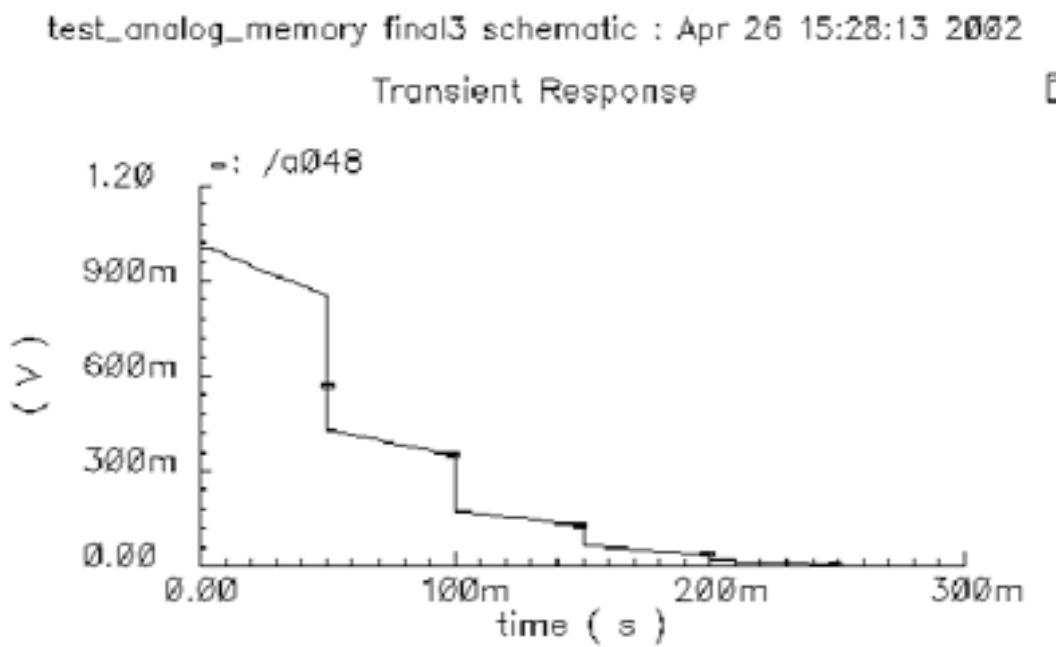

Figure 3.15  Voltage fall in $C_{primary}$

Above graph shows the voltage across $C_{primary}$. In the sharp transitions it had charge sharing with $C_{secondary}$ and then voltage dropped across it in usual rate.

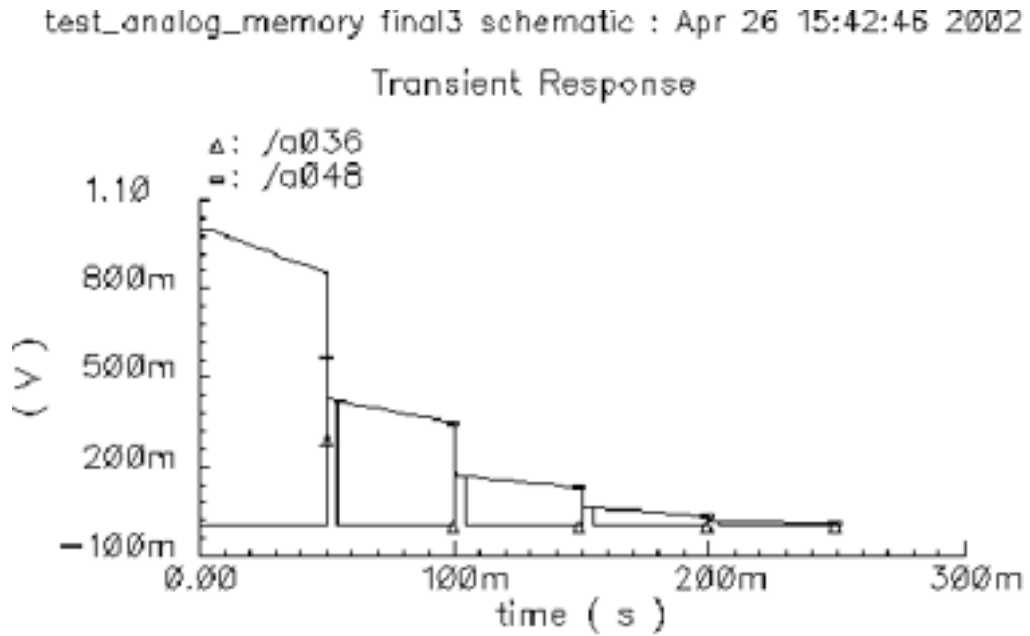

Figure 3.16 Comparison of voltages across primary and secondary capacitors

a036 : Voltage across $C_{secondary}$

a048 : Voltage across $C_{primary}$

From the value of voltage across the secondary capacitor we can conclude that it's amplification is required to get the correct analog value stored. So for reading purposes voltage gain is required and that is described in next chapter.

# Chapter 4

# Design of self refreshing Analog Memory Cell

## 4.1 Reading voltage stored from capacitor

### 4.1.1 Circuit

The part of the analog memory including reading the voltage stored is in fig. 4.1. The READ voltage out is taken from the *out* point of the op amp. The NMOS $M_3$ acts as the switch controlling the input of voltage into the op-amp. Whenever $V_{amplify}$ is HIGH, $M_3$ is ON and the voltage across $C_{secondary}$ comes to the non-inverting input of the op-amp. The op amp acts as a non-inverting amplifier where gain is given as

$$Gain = ( 1 + R_1 / R_0 )$$

i.e.     $V_{out} = ( 1 + R_1 / R_0 ) * V_{in}$.

The gain of the op amp was adjusted in such a way that the output voltage gives the initial voltage stored in the primary capacitor. In other words the gain of the op-amp is used to make up for the charge loss in both the capacitors and sharing of the charge between them.

The op amp amplifies the voltage stored across $C_{secondary}$. This amplified voltage at the point *out* then becomes the READ OUT signal from the analog memory cell.

Figure 4.1 Analog memory with writing to and reading from cell (no refreshing)

# 4.1.2      Simulation and Results

The resistances $R_1$ and $R_0$ are the key parameters in adjusting gain of the op-amp. Their values were decided in the following way. For various values of the input voltage from 0.2 to 2 volts the rate of fall was studied i.e. after required time of 40 ms how much the voltage has fallen to. This reduced voltage must be up converted by the op amp to the initial value stored. The analysis gives the amount of gain required by the op-amp and hence values of $R_1$ and $R_0$ can be found out.

The table 4.1 shows the results of the analysis and hence the value of $R_1$ required.

In the calculations took $R_0$ = 500 ohms and value of $R_1$ was found out. From Table 4.1 we can see that the required value of $R_1$ required is between 1.5835 K ohms and 1.742 K ohms for most accurate results.

In the project for sake of simplicity $R_1$ was approximated to be 1.7 K ohms. So the refreshed voltage was accurate as the input voltage within certain tolerance limits as shown in results next.

| Serial No. | Required Output of op amp ie. value of Vin $V_{OUTPUT}$ ( volts ) | Input to op amp ie. voltage across $C_{secondary}$ $V_{IN}$ ( volts ) | ( 1 + $R_1$ / $R_0$ ) GAIN | Taking $R_0 = 500$ ohms, Required $R_1$ ( K ohms ) |
|---|---|---|---|---|
| 1. | 0.2 | 48m | 4.167 | 1.5835 |
| 2. | 0.3 | 72m | 4.167 | 1.5835 |
| 3. | 0.4 | 95m | 4.21 | 1.605 |
| 4. | 0.5 | 0.118 | 4.24 | 1.62 |
| 5. | 0.6 | 0.14 | 4.286 | 1.643 |
| 6. | 0.7 | 0.162 | 4.32 | 1.66 |
| 7. | 0.8 | 0.185 | 4.324 | 1.662 |
| 8. | 0.9 | 0.206 | 4.369 | 1.6845 |
| 9. | 1.0 | 0.228 | 4.386 | 1.693 |
| 10. | 1.1 | 0.25 | 4.4 | 1.7 |
| 11. | 1.2 | 0.273 | 4.4 | 1.7 |
| 12. | 1.3 | 0.295 | 4.407 | 1.7035 |
| 13. | 1.4 | 0.32 | 4.41 | 1.705 |
| 14. | 1.5 | 0.338 | 4.438 | 1.719 |
| 15. | 1.6 | 0.36 | 4.44 | 1.72 |
| 16. | 1.7 | 0.383 | 4.44 | 1.72 |
| 17. | 1.8 | 0.402 | 4.477 | 1.7385 |
| 18. | 1.9 | 0.424 | 4.48 | 1.74 |
| 19. | 2.0 | 0.446 | 4.484 | 1.742 |

Table 4.1 Determining value of $R_1$ from Gain required

With the values of resistances as $R_0$ = 500 ohms and $R_1$ = 1.7 K ohms the following graphs were obtained. Here input voltage $V_{in}$ = 1 volts.

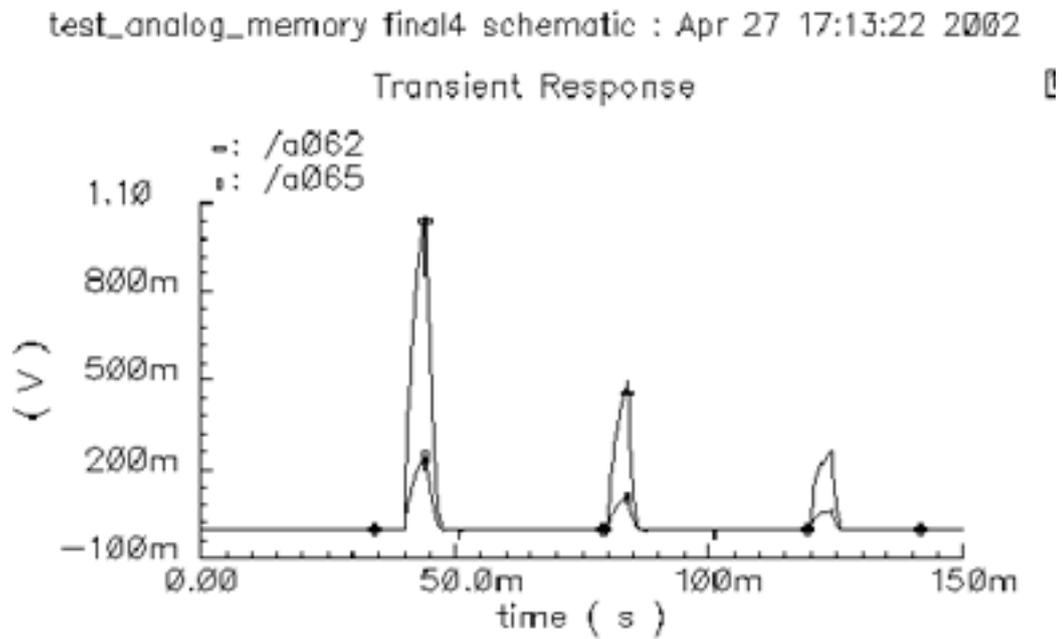

Figure 4.2 Input and Output voltages of op amp

a065:Input voltage to op amp

a062:Output voltage from op amp

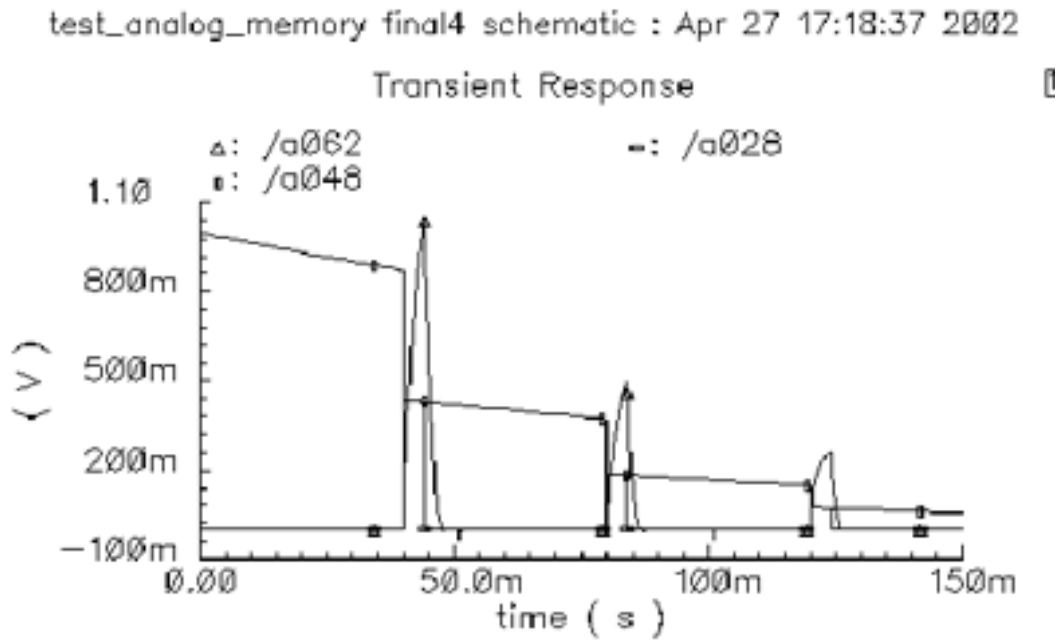

Figure 4.3   Voltages across primary and secondary capacitors as compared to the voltage amplified by the op amp

a048:voltage across primary capacitor

a028:voltage across secondary capacitor

a062:voltage amplified by the op amp

So here we conclude that refreshing of voltage of the primary capacitor is required to read voltage correctly from it.

## 4.2 Final analog memory cell with refreshing arrangement

### 4.2.1 Circuit

The input analog dc voltage to be stored is given as a narrow pulse, $V_{in}$. At the same time the control voltage $V_{write}$ makes the MOS switch $M_0$ ON and the input voltage is stored in $C_{primary}$. Then whenever the control voltage $V_{transfer}$ makes the MOS switch M1 ON; charge stored in $C_{primary}$ gets distributed between itself and $C_{secondary}$. When $V_{amplify}$ makes the MOS $M_3$ ON, voltage from $C_{secondary}$ goes to the input of the op amp and gets amplified at it's output. The values of resistances $R_0$ and $R_1$ adjusting the gain of the op amp were chosen from previous analysis of section 4.1. The amplified voltage from op amp is then used to refresh the primary capacitor by the switch $M_4$ controlled by the voltage $V_{refresh}$. Similar to other MOS switches, whenever $V_{refresh}$ at gate of $M_4$ is HIGH, $M_4$ switches ON and passes the voltage amplified by the op amp to the primary capacitor ; hence refreshing $C_{primary}$.

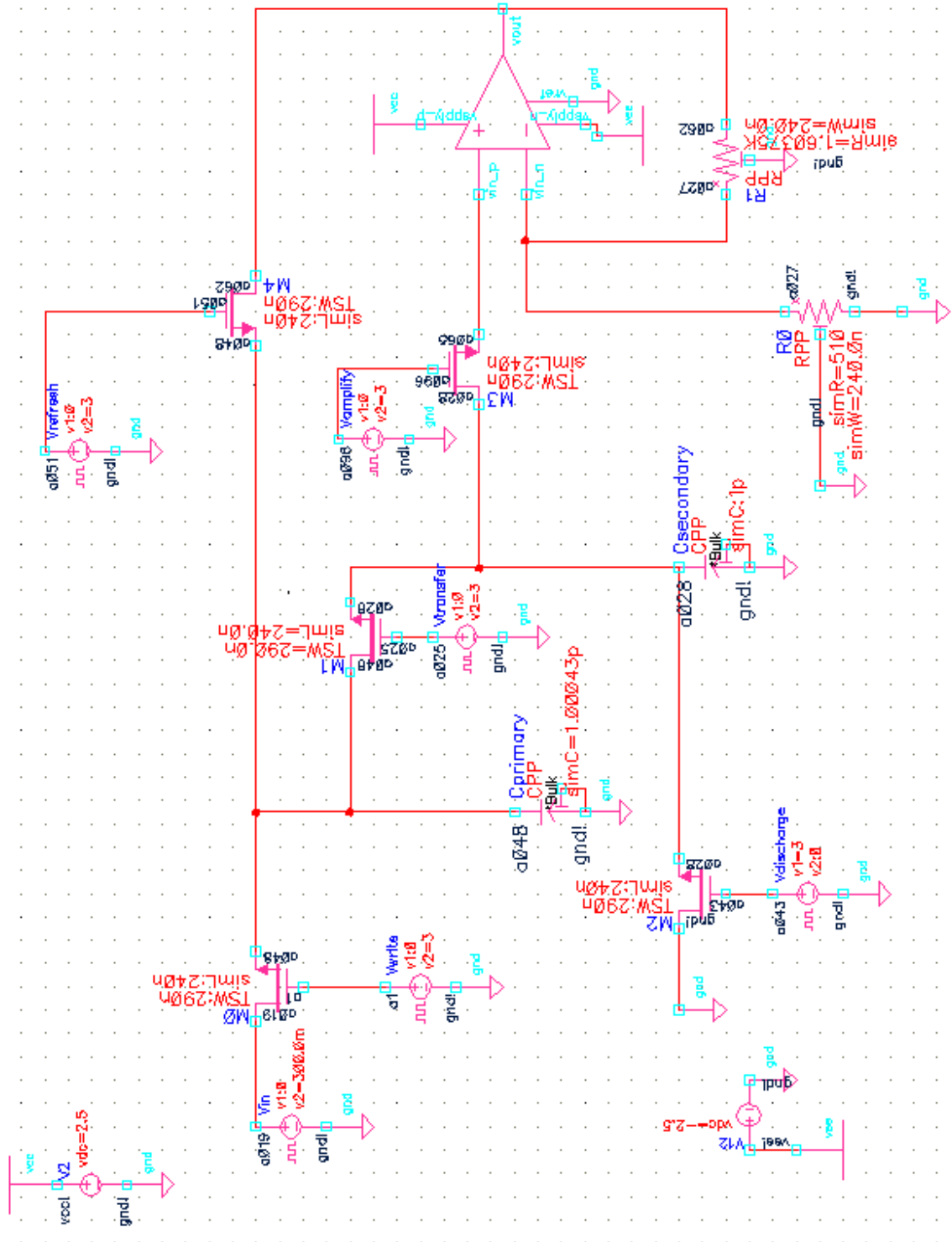

Figure 4.4    Complete Circuit of Analog Memory Cell
( with writing, reading and refreshing facility )

## 4.2.2 Simulation and Results

The following graphs show the transient responses at various points of the analog memory cell i.e. the plots of voltages with respect to time.

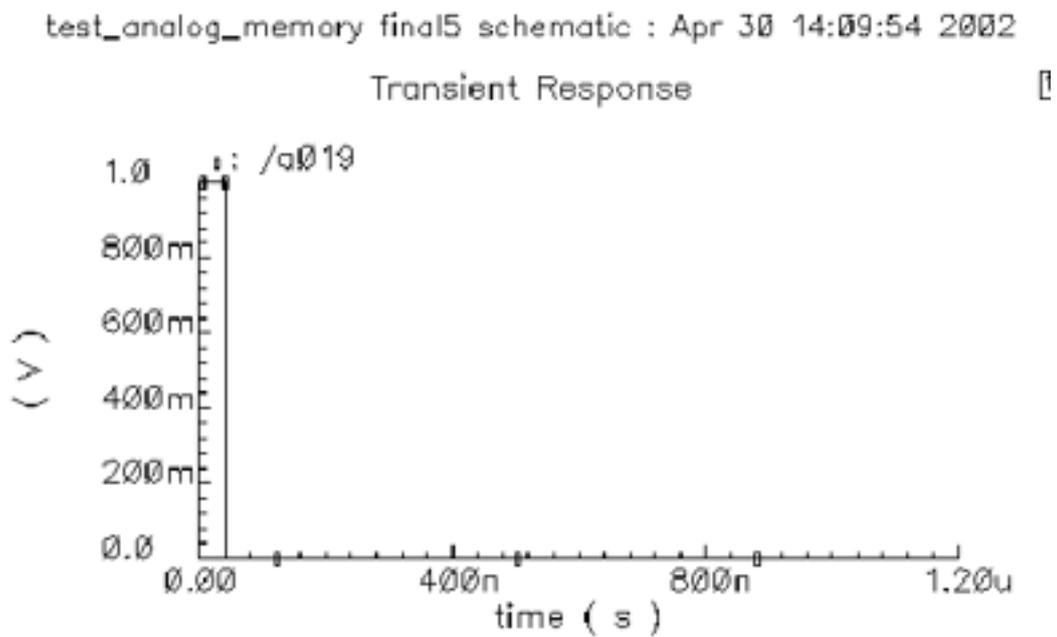

Figure 4.5   $V_{in}$, Input pulse given

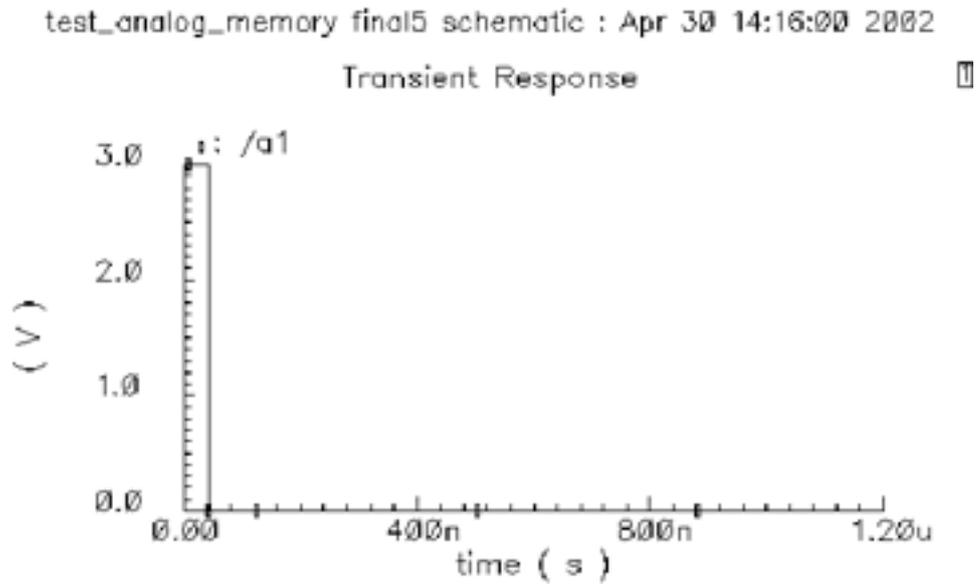

Figure 4.6    V_write, Pulse controlling writing of input voltage to primary capacitor

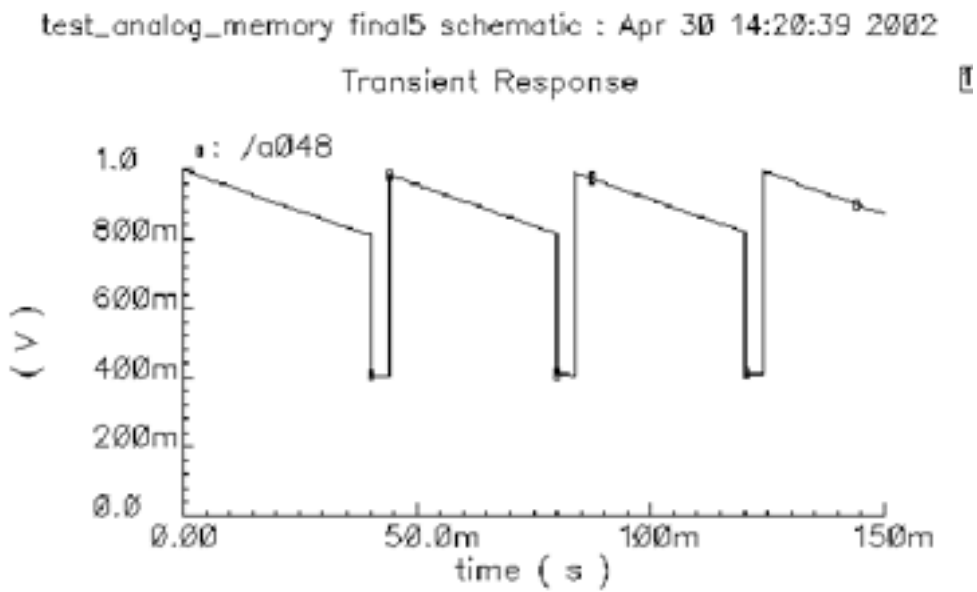

Figure 4.7    Voltage across the primary capacitor

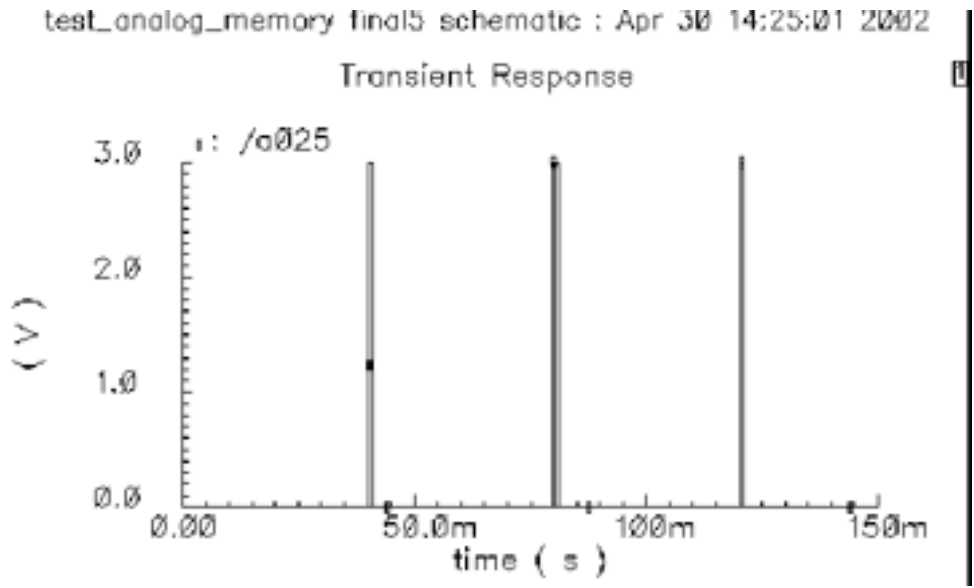

Figure 4.8  Control signal V$_{transfer}$ given to transfer charge from primary to secondary capacitor

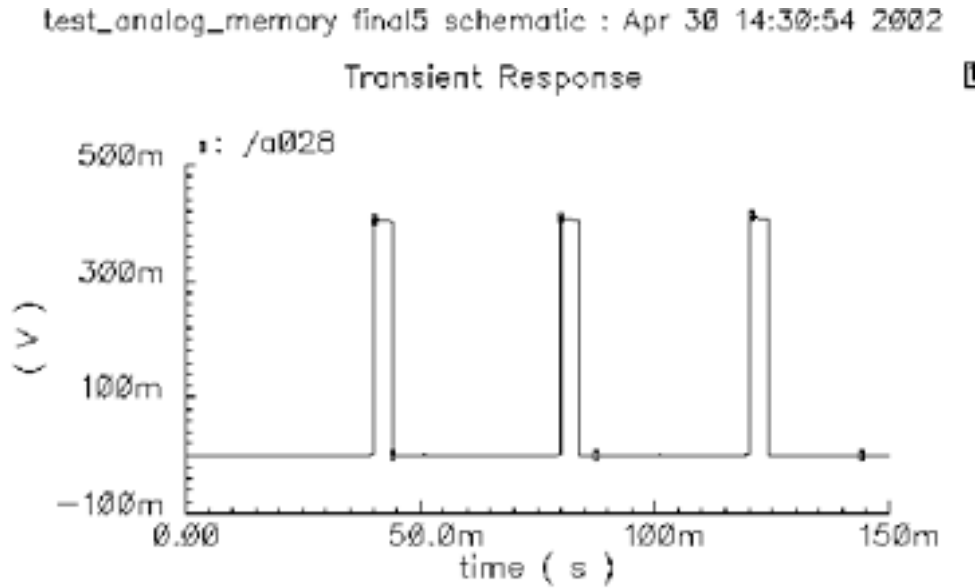

Figure 4.9  Voltage across the secondary capacitor

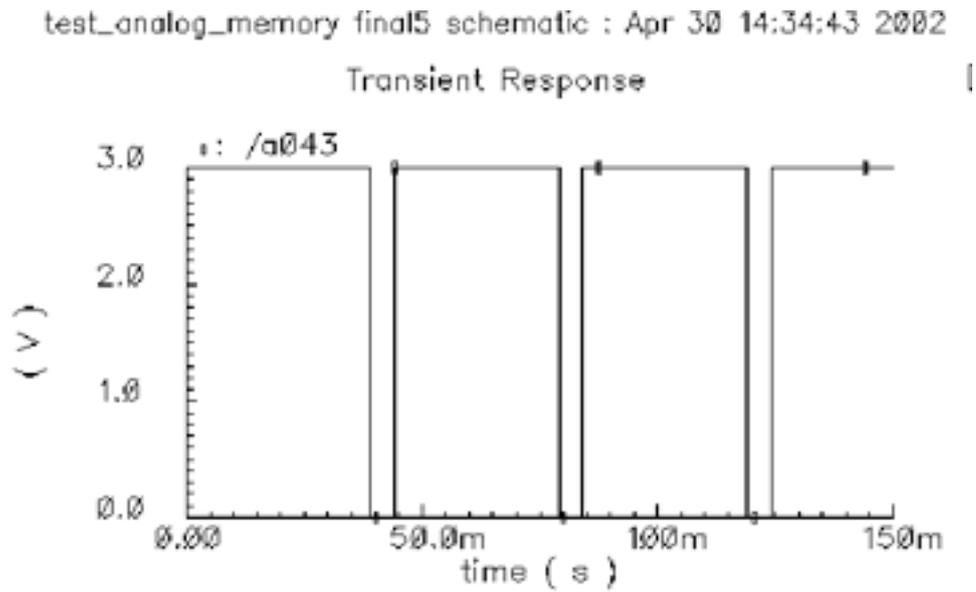

Figure 4.10  Voltage V<sub>discharge</sub> given to discharge secondary capacitor when charge is not stored in it

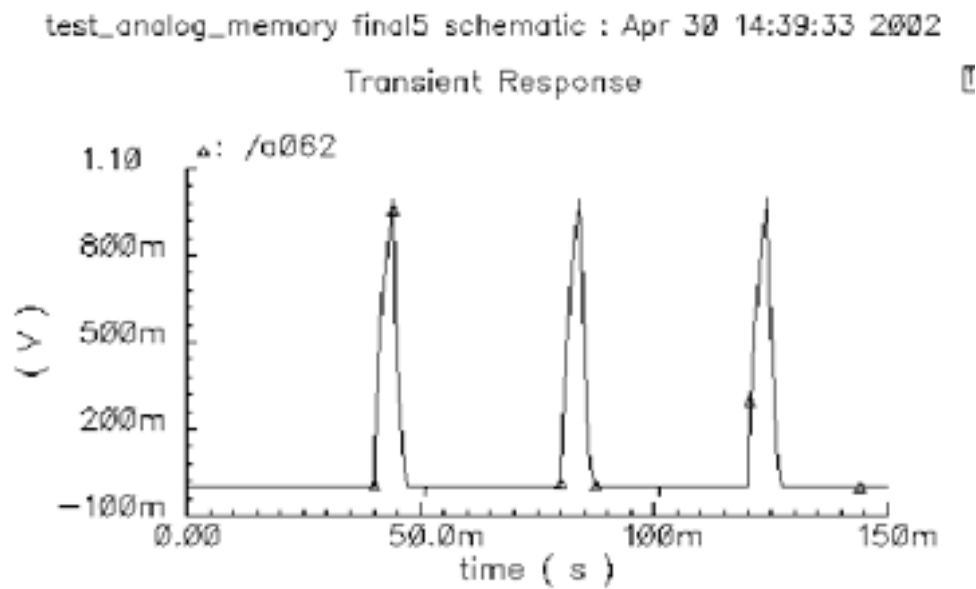

Figure 4.11  Amplified voltage from op amp i.e. the voltage read out

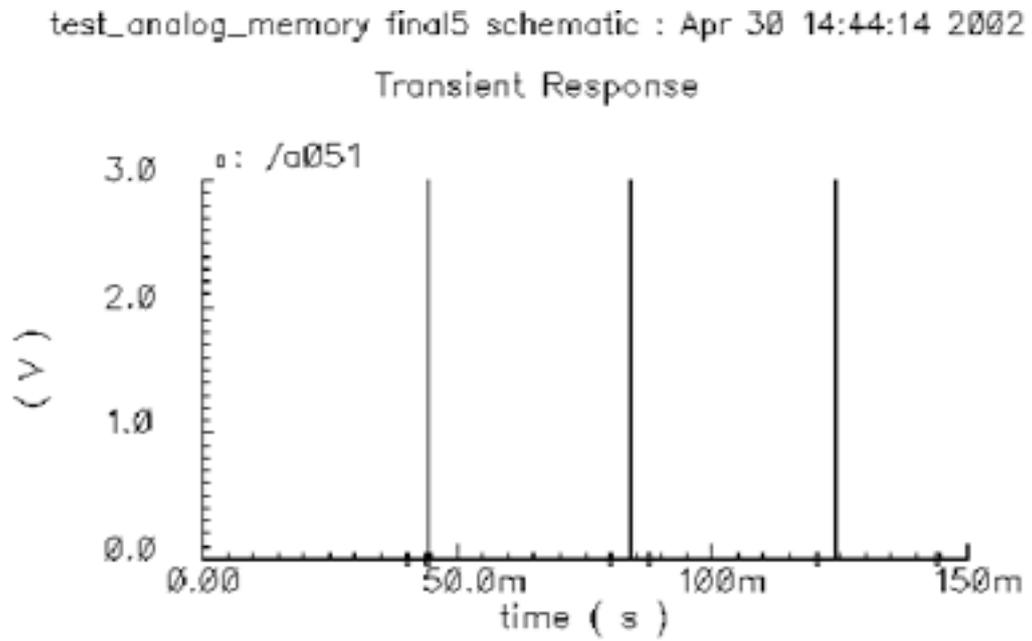

Figure 4.12  Control voltage V<sub>refresh</sub>

The following graphs give a comparison of the quality of the analog memory cell when storing various amounts of input voltage. The resistance R1 was chosen to be 1.7 K ohms, giving accurate gain for input voltages around 1volts. Hence as in fig. 4.13, for input voltage of 0.4 volts after 120 milliseconds the voltage is 0.42 volts.

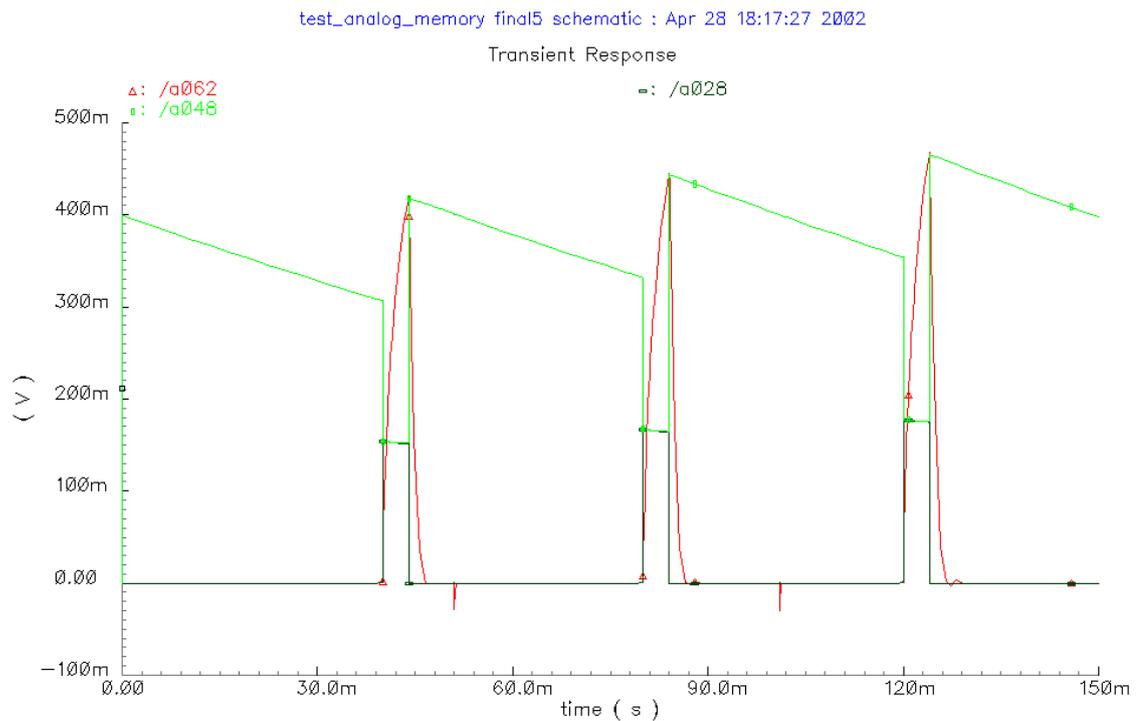

Figure 4.13  Comparison of Various voltages when $V_{in}$ = 0.4 volts.

       a048  :     voltage across primary capacitor

       a028  :     voltage across secondary capacitor

       a062  :     voltage amplified by the op amp

From fig. 4.14, input voltage of 1 volt after 120 milliseconds of storage remains at 1 volt.

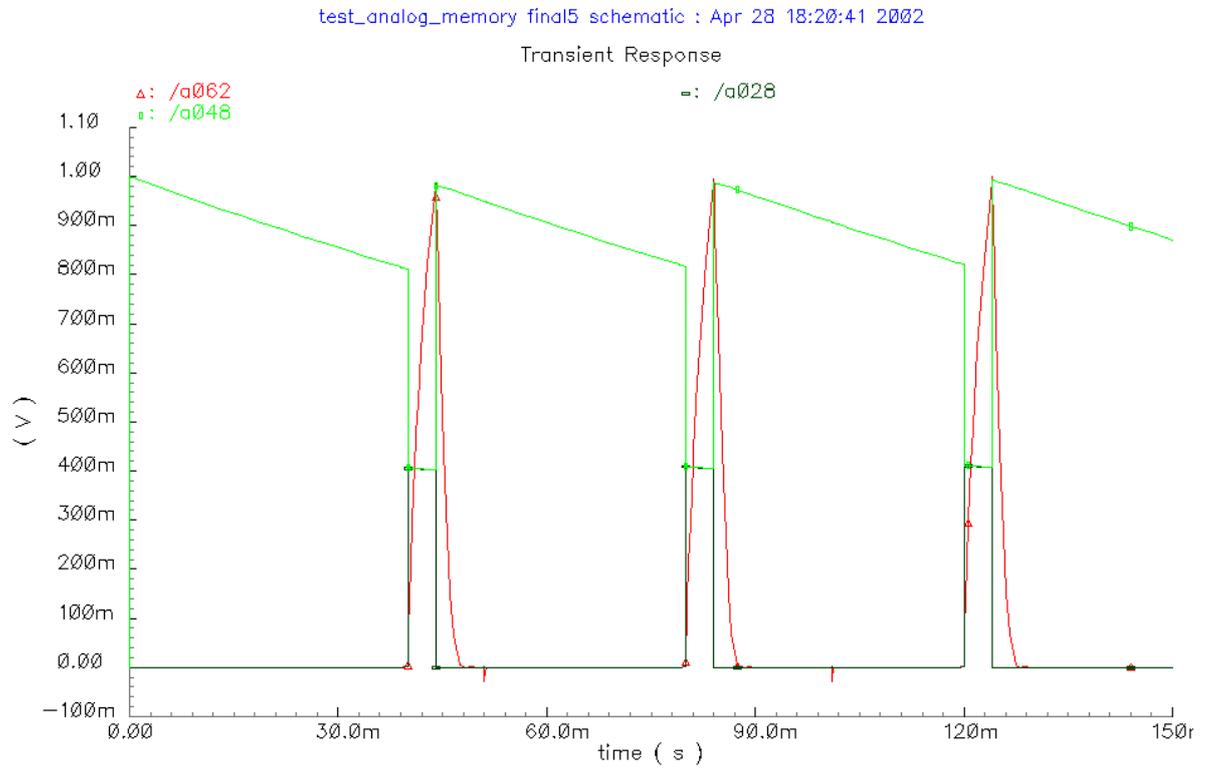

Figure 4.14  Comparison of Various voltages when $V_{in}$ = 1 volts.

a048  :  voltage across primary capacitor

a028  :  voltage across secondary capacitor

a062  :  voltage amplified by the op amp

From fig. 4.15, input analog voltage of 1.9 volt after 120 milliseconds of storage becomes 1.85 volts.

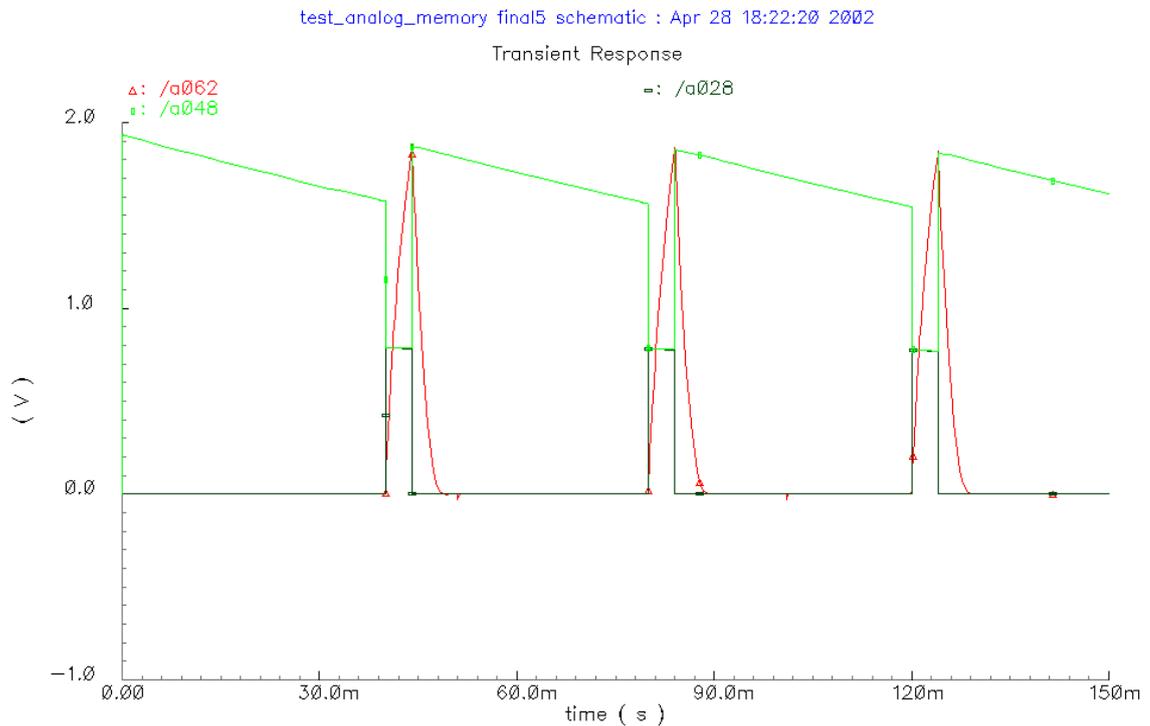

Figure 4.15  Comparison of Various voltages when V<sub>in</sub> = 1.9 volts.

        a048  :      voltage across primary capacitor

        a028  :      voltage across secondary capacitor

        a062  :      voltage amplified by the op amp

## _Results :_

1. All the next results are based on the assumptions that voltage is to be stored till 120 milliseconds, and charge is to be refreshed in primary capacitor every 40 milliseconds assuming frame rate is 25 frames per second.

| Serial No. | Input analog voltage $V_{IN}$ | Voltage after 120 ms $V_{LATER}$ | Error appeared in storage (as % of $V_{IN}$) |
|---|---|---|---|
| 1. | 0.2 | 0.23 | 15 |
| 2. | 0.3 | 0.33 | 10 |
| 3. | 0.4 | 0.42 | 5 |
| 4. | 0.5 | 0.52 | 4 |
| 5. | 0.6 | 0.61 | 1.67 |
| 6. | 0.7 | 0.71 | 1.43 |
| 7. | 0.8 | 0.8 | 0 |
| 8. | 0.9 | 0.9 | 0 |
| 9. | 1.0 | 1.0 | 0 |
| 10. | 1.1 | 1.1 | 0 |
| 11. | 1.2 | 1.2 | 0 |
| 12. | 1.3 | 1.3 | 0 |
| 13. | 1.4 | 1.4 | 0 |
| 14. | 1.5 | 1.48 | 1.33 |
| 15. | 1.6 | 1.58 | 1.25 |
| 16. | 1.7 | 1.68 | 1.18 |
| 17. | 1.8 | 1.76 | 2.22 |
| 18. | 1.9 | 1.85 | 2.63 |
| 19. | 2.0 | 1.94 | 3 |

Table 4.2 Quality of storage of analog voltage in the memory (as % error)

2. The analog memory cell stored voltages with accuracy as shown in Table 2. From there it can be concluded that the memory works very good for storing voltage in the range 0.4 volts to 2 volts with maximum 5 % error. It can go till 0.2 volts in lower range where error is somewhat more, around 15 %. If analog voltage to be stored decreases beyond 0.2 volts, error crept up increases considerably.

   So the circuit is said to store analog voltage in the range 0.2 volts to 2 volts.

3. In the analog memory cell designed, the refreshing interval is to be decided depending on the frame rate. If frame rate is different from 25 frames per second, refreshing interval is to be chosen appropriately so that before the next frame comes the value of previous frame is refreshed and read accurately. This can be done easily using a similar kind of analysis for gain of op amp, as done in the project.

4. The input analog voltage must be present till a time greater than 40 nanoseconds, and the write control voltage, $V_{write}$ must be present till at least 40 nanoseconds for analog voltage to be written successfully in the capacitor. So the highest speed with which it can work is 25 MHz.

5. Voltage is stored constant within 5 % accuracy till 120 milliseconds. So it can work with slow systems as much as 8 frames per second.

# 4.3 Layout

## 4.3.1 Circuit Schematic for layout

The next circuit shows the modification to be done in schematic for doing the layout.

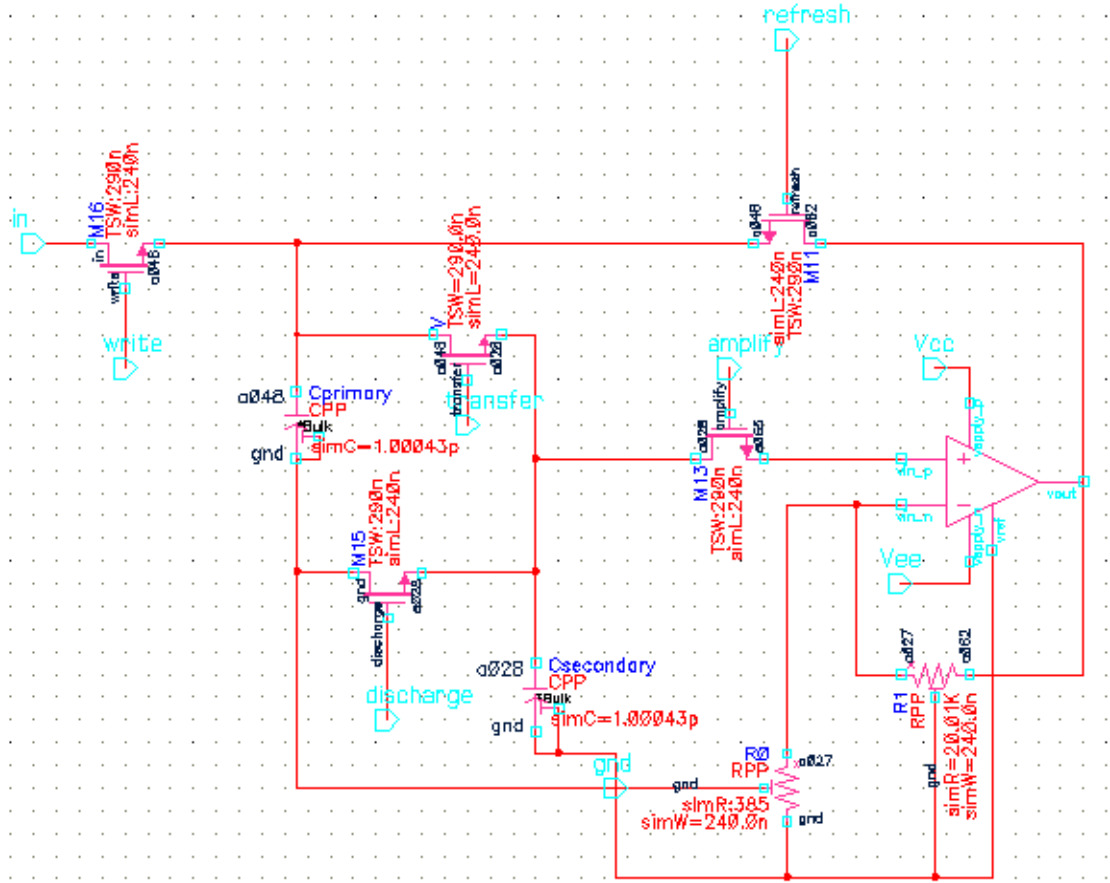

Figure 4.16  Schematic of Analog Memory Cell for Layout

## 4.3.2 Layout ( Partial ) of the Analog Memory Cell

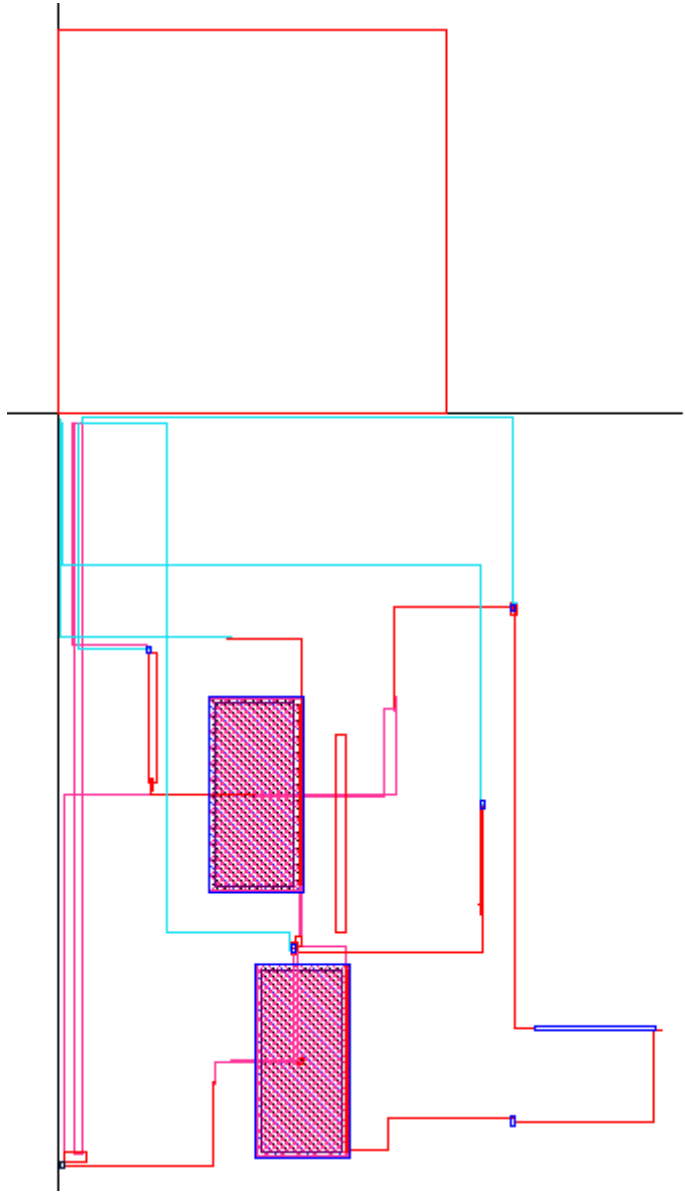

Figure 4.17  Layout of the designed Analog Memory Cell

# Chapter 5

# Conclusions

In the project at first a detailed review of the literature available and previous works in Analog Memory was done. The various alternative design techniques for analog memory like switched capacitor method or switched current method were reviewed.

A new analog memory cell was developed using the switched capacitor technique.

The memory cell can store analog voltage values with 5 % accuracy in the input range 0.4 volts to 2 volts. It can go till 0.2 volts but then error goes up to 15 %. If it goes further below 0.2 volts error in storage increases considerably.

Practical applications of the analog memory designed is in large scale Consumer Market, where quality and cost both are important. Generally common people try to optimize cost vs. quality. They look for low cost and the best quality possible in it. Analog memory is a very good way to reduce cost of electronic goods but not losing quality so much.

Analog memory can be used in systems like Videophone. A hi tech system with accurate digital memory will cost very high. But if analog memory is used, quality or accuracy of storage may get halved with as much reduction in cost as one tenth. In that case people will go for design with analog memory as it gives them more optimization.

# **Future Scopes**

- In the circuit a linear op amp is used which has a constant gain adjusted by the two constant biasing resistances $R_1$ and $R_0$. But by previous analysis it was seen that if $R_1$ varies between 1.5 K ohms and 1.7 K ohms then we get the best results. This is because then gain of op amp is such adjusted that for varying amounts of input voltage the output is same as the analog voltage IN.

  So if a non-linear op amp is used with variable amounts of gain, instead of the linear one, results will be better. This can be done by using a non-linear resistance in place of $R_1$.

- In the project a dual power supply of + 2.5 volts and – 2.5 volts was used. This supply voltage was required for only one component of the circuit, the op amp. In the CMOS technology used the op amp was taken from the library available, and it required a dual power supply. It was a typical op amp in non-inverting mode.

  This drawback of using dual power supply can be removed by proper circuit design techniques. If rail-to-rail op amp is used then a single unipolar power supply of 0 to + 2.5 volts will do. But then the technology used will have to be changed. This can be done by future studies and detailed analysis.

# Appendix A

# CMOS Device Modeling for Circuit Design

Before a circuit is designed, to be integrated by CMOS VLSI technology, a model must be adopted which will describe behavior of all components successfully. A model means a set of mathematical formulas, circuit representations, tables, reference standards etc.

## A.1. Simple MOS large signal model

Lower case variables represent instantaneous values and upper case variables represent averaged or rms values. Lower case subscripts are used for small signal - values and upper case subscripts for large – signal values. This model was developed by Sah and is also called the SPICE Level 1 Model. This model is given by

$$i_D = \frac{\mu_0 C_{ox} W}{L} \left[ (v_{GS} - V_T) - \left( \frac{v_{DS}}{2} \right) \right] v_{DS} \left( 1 + \lambda v_{DS} \right) \qquad (\text{for } v_{DS} < v_{GS} - V_T) \quad (A.1)$$

$$i_D = \frac{\mu_0 C_{ox} W}{2L} \left(v_{GS} - V_T\right)^2 \left(1 + \lambda v_{DS}\right) \qquad \text{(for } v_{DS} > v_{GS} - V_T\text{)} \qquad \text{(A.2)}$$

$\mu_0$ = surface mobility of the channel for device (cm$^2$/V-s)

$C_{ox} = \frac{\epsilon_{ox}}{t_{ox}}$ = capacitance per unit area of the gate oxide (F/cm$^2$)

$W$ = effective channel width

$L$ = effective channel length

$$V_T = V_{T0} + \gamma \left[\sqrt{2|\phi_F| + v_{SB}} - \sqrt{2|\phi_F|}\right] \qquad \text{(A.3)}$$

$$V_{T0} = V_{T(v_{SB}=0)} = V_{FB} + 2|\phi_F| + \frac{\sqrt{2q\epsilon_{si}N_{SUB}2|\phi_F|}}{C_{ox}} \qquad \text{(A.4)}$$

$$\gamma = \text{bulk threshold parameter}\left(V^{1/2}\right) = \frac{\sqrt{2\epsilon_{si}q N_{SUB}}}{C_{ox}} \qquad \text{(A.5)}$$

$$\phi_F = \text{strong inversion surface potential (V)} = -\frac{kT}{q}\ln\left(\frac{N_{SUB}}{n_i}\right) \qquad \text{(A.6)}$$

$$V_{FB} = \text{flatband voltage (V)} = \phi_{MS} - \frac{Q_{ss}}{C_{ox}} \qquad \text{(A.7)}$$

$$\phi_{MS} = \phi_F(\text{substrate}) - \phi_F(\text{gate}) \qquad \text{(A.8)}$$

$$\phi_F(\text{substrate}) = -\frac{kT}{q}\ln\left(\frac{N_{SUB}}{n_i}\right) \qquad \text{(A.9)}$$

$$\phi_F(\text{gate}) = -\frac{kT}{q} \ln\left(\frac{N_{GATE}}{n_i}\right) \tag{A.10}$$

$$Q_{ss} = \text{oxide charge} = qN_{ss} \tag{A.11}$$

$k$ = Boltzmann's constant

$T$ = temperature (K)

$n_i$ = intrinsic carrier concentration

Table A.1 gives the pertinent constants for silicon. A unique aspect of the MOS device is its dependence upon the voltage from the source to bulk as shown by equation A.3.This dependence means that the MOS device must be treated as a four-terminal element. The above equations are valid in the reqion $v_{DS} < v_{GS} - V_T$, called the linear region.

| Constant Symbol | Constant Description | Value | Units |
|---|---|---|---|
| $V_G$ | Silicon Bandgap $(27^\circ C)$ | 1.205 | V |
| $k$ | Boltzmann's constant | $1.381 \times 10^{-23}$ | J/K |
| $n_i$ | Intrinsic carrier concentration $(27^\circ C)$ | $1.45 \times 10^{10}$ | $cm^{-3}$ |
| $\epsilon_0$ | Permittivity of free space | $8.854 \times 10^{-14}$ | F/cm |
| $\epsilon_{si}$ | Permittivity of silicon | $11.7\epsilon_0$ | F/cm |
| $\epsilon_{ox}$ | Permittivity of SiO$_2$ | $3.9\epsilon_0$ | F/cm |

Table A.1: Constants for Silicon

More detailed models for $i_D$ and other large signal parameters such as capacitances can be found in Sze [1], Allen & Holberg [2], Laker & Sansen [3] or Gray & Meyer [4].

Another important parameter for circuit performance is noise. The existence of noise is due to the fact that electrical charge is not continuous but is carried in discrete amounts equal to the charge of an electron. In electronic circuits, noise manifests itself by representing a lower limit below which electrical signals cannot be amplified without significant deterioration in the quality of the signal. Noise can be modeled by a current source connected in parallel with $i_D$. This current source represents two source of noise, called thermal noise and flicker noise [2][4]. More advanced noise models can be found in [6] and references therein. The mean-square current-noise source is defined as

$$\bar{i}_N^2 = \left[ \frac{8kTg_m(1+\eta)}{3} + \frac{(KF)I_D}{fC_{ox}L62} \right] \Delta f \tag{A.12}$$

Reflected to the input, this becomes

$$\bar{i}_{eq}^2 = \left[ \frac{8kT(1+\eta)}{3g_m} + \frac{KF}{2fC_{ox}WLK\prime} \right] \tag{A.13}$$

where

$$
\begin{aligned}
\Delta f &= \quad \text{bandwidth at a frequency } f \\
\eta &= \quad g_{mbs}/g_m \\
k &= \quad \text{Boltzmann's constant} \\
T &= \quad \text{temperature (K)} \\
g_m &= \quad \text{small-signal transconductance from gate to channel} \\
KF &= \quad \text{flicker-noise coefficient(F-A)} \\
f &= \quad \text{frequency (Hz)}
\end{aligned}
$$

## A.2. Simple MOS Small – Signal Model

For analog circuit design a few equations are most commonly used and must be remembered. These equations can give the dc bias conditions and low frequency characteristics. However, the ac analysis can be done using more involved expressions [2][4][3] not discussed in this thesis.

$$i_D = \beta \left[ (v_{GS} - V_T) - \frac{v_{DS}}{2} \right] v_{DS} \qquad \text{(in linear region)} \qquad \text{(A.14)}$$

$$i_D = \frac{\beta}{2} (v_{GS} - V_T)^2 \qquad \text{(in saturation)} \qquad \text{(A.15)}$$

$$v_{DS(sat)} = \sqrt{\frac{i_D}{2\beta}} \qquad \text{(A.16)}$$

$$g_m = \text{transconductance} = \sqrt{2\beta i_D} \qquad \text{(A.17)}$$

$$r_{ds} = \text{small signal drain resistance} = \frac{1}{\lambda I_D} \qquad \text{(A.18)}$$

# Appendix B

# The Cadence IC 445 set of tools

The IC 4.4.5 set of tools is a complete front-end to back-end design suite for integrated circuit design. This chapter gives a summary of the different design tools available to an analog circuit designer.

## B.1  Design Entry

For analog design, the Composer Schematic Editor tool is used. This allows easy entry of the schematic. The editor employs several useful options to make design entry efficient. Some of these include gravity, object sensitive dialogs and pop-up menus. It also includes a method for checking the validity of the design. The schematic entry environment extracts a netlist called the CDL.

## B.2  Simulation Environment

The analog artist simulation environment is used to simulate the design. This is a front-end to the actual simulator. It allows one to choose the simulator and set the 54 options appropriate for the simulator through easy to use menus. Analog artist also includes extensive post-processing facilities for analyzing the results of the circuit simulation. Analog artist contains support for cdsSPICE, spectreS, HSPICE and other simulators. cdsSPICE and spectreS are Cadence's simulators based on SPICE but with some extensions. It supports all standard SPICE models. Specifically, BSIM3v3 models were used for the simulation of the circuits.

## B.3  Design Synthesis and Layout

The device level editor (DLE) is used to generate the layout from the schematic. This places all the devices in a layout window approximately in accordance with their placement in the schematic. This also has facility for showing the devices and nets that need to be connected. This allows one to do a layout that corresponds exactly to the schematic. The Virtuoso layout editor is used for defining the interconnects and other structures in the layout.

## B.4  Design Verification

Once the layout is made, it has to be verified to be correct. For this, one does a design rule check (DRC) using the Diva verification tool. After the design rule check has been passed, a netlist has to be extracted from the layout. This netlist is then compared with the schematic netlist using the "Layout vs. Schematic" option in Diva. Diva also contains options for extraction of parasitic resistance and capacitance. This information can be back annotated into the circuit and the simulation repeated to verify that the circuit does indeed work with the parasitics.

# Appendix C

# Acronyms

**ADC**　　　　Analog to Digital Converter

**BiCMOS**　　Bipolar Complementary Metal Oxide Semiconductor

**BJT**　　　　Bipolar Junction Transistor

**CAD**　　　　Computer Aided Design

**CMOS**　　　Complementary Metal Oxide Semiconductor

**DAC**　　　　Digital to Analog Converter

**IC**　　　　　Integrated Circuit

**MOS**　　　　Metal Oxide Semiconductor

**MOSFET**　　Metal Oxide Semiconductor Field Effect Transistor

**VLSI**　　　Very Large Scale Integration